\documentclass[fleqn,10pt]{wlscirep}
\usepackage[utf8]{inputenc}
\usepackage[T1]{fontenc}
\usepackage{xurl}   
\usepackage{graphicx}
\usepackage{dcolumn}

\usepackage{bm}
\usepackage{romannum}
\usepackage{amsmath}
\usepackage{graphicx}
\usepackage{xcolor}
\usepackage[caption=false]{subfig}
\usepackage{booktabs}   
\usepackage{relsize}

\usepackage{float}

\usepackage[justification=justified,singlelinecheck=false]{caption}

\restylefloat{table}

\title{A smooth filament origin for distant prolate galaxies seen by JWST and HST}

\author[1,*]{Alvaro Pozo}
\author[1,2,3]{Tom Broadhurst}
\author[4]{Razieh Emami}
\author[5]{Philip Mocz}
\author[6]{Mark Vogelsberger}
\author[4]{Lars Hernquist}
\author[7]{Christopher J. Conselice}
\author[1]{Hoang Nhan Luu}
\author[1,8,9,10]{George F. Smoot}
\author[11]{Rogier Windhorst}

\affil[1]{DIPC, Basque Country UPV/EHU,San Sebastian, E-48080, Spain}

\affil[2]{University of the Basque Country UPV/EHU, Department of Theoretical Physics, Bilbao, E-48080, Spain}

\affil[3]{Ikerbasque, Basque Foundation for Science, Bilbao, E-48011, Spain}

\affil[4]{Center for Astrophysics $\vert$ Harvard \& Smithsonian, 60 Garden Street, Cambridge, MA 02138, USA}

\affil[5]{Center for Computational Astrophysics, Flatiron Institute, 162 5th Ave., New York, NY, 10010, USA}

\affil[6]{Dept. of Physics, Kavli Institute for Astrophysics \& Space Research, Massachusetts Institute of Technology, Cambridge, MA 02139, USA}

\affil[7]{Jodrell Bank Centre for Astrophysics, University of Manchester, Oxford Road, Manchester, UK}

\affil[8]{Department of Physics and Institute for Advanced Study, The Hong Kong University of Science and Technology, Hong Kong}

\affil[9]{Paris Centre for Cosmological Physics, APC, AstroParticule et Cosmologie, Universit´e de Paris, CNRS/IN2P3, CEA/lrfu, 10, \\ rue Alice Domon et Leonie Duquet, 75205 Paris CEDEX 13, France emeritus}

\affil[10]{Physics Department, University of California at Berkeley, CA 94720, Emeritus}

\affil[11]{School of Earth and Space Exploration, Arizona State University, Tempe, AZ 85287-6004, USA}

\affil[*]{alvaro.pozolarrocha@bizkaia.eu}
\begin{document}


\begin{abstract}
\textbf{The initial gravitational collapse of Dark Matter and gas forms a universal filamentary network where
the first galaxies form, with shapes and sizes that depend on the choice of Dark Matter. Claims from deep space imaging surveys that elongated galaxies predominate at $z > 3$ are examined here by comparison with detailed hydrodynamical simulations of Cold Dark Matter (CDM), Warm Dark Matter (WDM), and Wave/Fuzzy Dark Matter, $\psi$DM. For CDM and WDM we have sufficient volume, $10^{3}\,\mathrm{Mpc/h}^{3}$, to generate galaxies with stellar masses $> 10^{9}\,M_{\odot}$ at $z > 2$, allowing comparison with the CEERS and CANDELS surveys. We find the observed tendency towards elongated, prolate-shaped young galaxies is well matched by WDM, from material accreted along smooth filaments during the first $\simeq 500\,\mathrm{Myr}$, with little dependence on stellar mass. This contrasts with CDM, where the stellar morphology is mainly spheroidal, formed from merging of fragmented filaments. For CDM, several subhalos are predicted to be visible, whereas for WDM and $\psi$DM, early merging is rare. Our findings show how the shapes and sizes of early galaxies are sensitive to the smoothness of the underlying filament network, providing a new constraint on the nature of dark matter.}
\end{abstract}

\flushbottom
\maketitle

%
%
\thispagestyle{empty}

\section{Introduction} \label{sec:intro}

The outstanding capabilities of the James Webb Space Telescope (JWST) allow young galaxies at very early times to be compared directly with the optical rest-frame morphologies of nearby galaxies. This is revising our understanding of the evolution of the luminosity function\cite{Fink,Oguri,LF3,Nathan} and the nature of galaxies at unprecedentedly high redshifts\cite{hz1,hz2,hz3,hz4}. It has also revealed a new widespread population of compact, Balmer-break galaxies above $z>4$ \cite{Labbe:2023,Mathee,Donnan,Kocevski,Kokorev,Wil_dot,Epoch} which show common broad line emission \cite{Naidu,Goulding,Geris} and stellar rotation in deep spectroscopy\cite{RotationJWST,DE}. Examples of unusually massive spheroids have been uncovered, as well as disk galaxies displaying spiral\cite{spiral2,Exception1} and bar morphology\cite{Bar1,Bar2} out to $z\simeq 5$, challenging standard LCDM growth predictions \cite{Car1, Exception0,exception3,Chal,Car2}. Additionally, the JWST/CEERS survey \cite{Pandya:2024,Finkelstein:2023} reveals evolution towards elongated galaxy morphology, indicating a dominance of prolate galaxies at z$>$3 when compared to the statistical projection of triaxial ellipsoids spanning the full range of spheroidal, oblate and prolate shapes. The majority of these galaxies are found to trace the projected "banana" shaped distribution of size versus projected ellipticity that is characteristic of prolate galaxies, rather than oblate disks. This distribution is also identified in the JWST/EPOCHS multi-survey analysis\cite{Excess}, extending to lower stellar mass and higher redshift, z$>$6.5. Prolate galaxies may also account for the high proportion of flattened\cite{Gibson:2024} and apparently disk-like galaxies\cite{Ferreira:2022} reported in deeper JWST imaging.

Early Hubble Space Telescope (HST) identifications of elongated ``chain" galaxies showing aligned HII regions in rest-UV images \cite{Cowie:1995,Elmgreen} were generally regarded as edge-on, given the geometric bias for optically thin disks. Subsequently, however, the case for intrinsically elongated high-z galaxies was reinforced in deeper Hubble images from the CANDELS survey and the Hubble Ultra-Deep Field (HUDF), where elongated, ``tadpole" galaxies were recognised\cite{Straughn:2006} and designated a common prolate class\cite{VDW:2014, Zhang:2019}. The survey of Ref\cite{Law:2012}, which provided spectroscopic redshifts for galaxies at 1.5$<$z$<$3.6, further found a preference for prolate triaxiality, with mean axis ratios $c/a=0.3$, $b/a=0.7$, over oblate disks.

This dominance of elongated galaxies at z$>$3 appears at odds with standard expectations conditioned by CDM simulations and the long-held notion that disks should form promptly when galaxies first collapse. If the majority of young galaxies at high redshifts are initially prolate, this implies that disk formation emerges later; however, disk evolution has been found to be relatively modest in HST-based morphological studies below z$<$1.5\cite{Odewahn:1997,Rogier1}, and disk galaxies have been securely identified by JWST at cosmic noon and beyond \cite{DISK1,Chal,DISK3,DISK4}.

Here we compare the stellar predictions for three contending DM scenarios -- Cold Dark Matter (CDM), Warm Dark Matter (WDM) and Wave/Fuzzy Dark Matter, $\psi$DM -- with the CEERS sample from Ref\cite{Pandya:2024} and the CANDELS sample from Ref\cite{VDW:2014}. We determine the 3D shapes of our simulated galaxies from their stellar morphology in the same way as Ref\cite{Pandya:2024}, and examine the trend with stellar age of each galaxy in our simulations and the shape evolution vs. stellar mass. We also predict the appearance of filaments at high-z and internal gas dynamics as further tests of our conclusions. A full set of simulated galaxies from three different simulations is presented here, with some previously published in Ref\cite{Mocz:2020} and conducted prior to the advent of JWST, thus representing a priori predictions.

\section{Results}


We now compare our simulations with CEERS and CANDELS in the redshift region $2<z<8$, where galaxies were found to be predominantly prolate in Ref\cite{Pandya:2024}. Our main comparison is with our CDM and WDM simulations for a 10~Mpc/$h$ box; two additional simulations are presented in the Supplementary Material, and the characteristics of all simulations are detailed in the Methods section. Both CDM and WDM were simulated with identical setups and resolution, and the outputs were analyzed in the same way. To ensure a fair comparison, we examined the exact same halos across both runs. For this purpose, once halos were identified by locating the particle with the minimum gravitational potential, we verified that their ID numbers and positions matched between the two simulations, guaranteeing a like-for-like comparison. When comparing 3D shapes, each datapoint corresponds to an individual galaxy. In contrast, when comparing 2D shapes, each data point represents one of the three orthogonal projections of a given galaxy, so that each galaxy contributes three data points. For the 10~Mpc/$h$ simulation, this corresponds to 2 individual halos at $z = 6$ (6 data points) and 25 halos at $z = 2$ (75 data points). Our simulated galaxies are matched in stellar mass and redshift to those from the CEERS and CANDELS datasets.

Extended Data Figure 1 shows the number of individual halos (and their corresponding orthogonal projections) analyzed at each redshift. Yellow bars indicate the halos used for the analysis at $z \geq 3$, while purple bars correspond to the halos analyzed at $z \leq 3$ (note that the $z = 3$ halos are included in both analyses). Since each halo is shown with its three orthogonal projections for 2D comparisons, the total number of data points is therefore three times the number of halos; these totals are represented by the transparent bars in Extended Data Figure 1.

\begin{figure}[H]
	\centering
	\includegraphics[width=180mm,height=100mm]{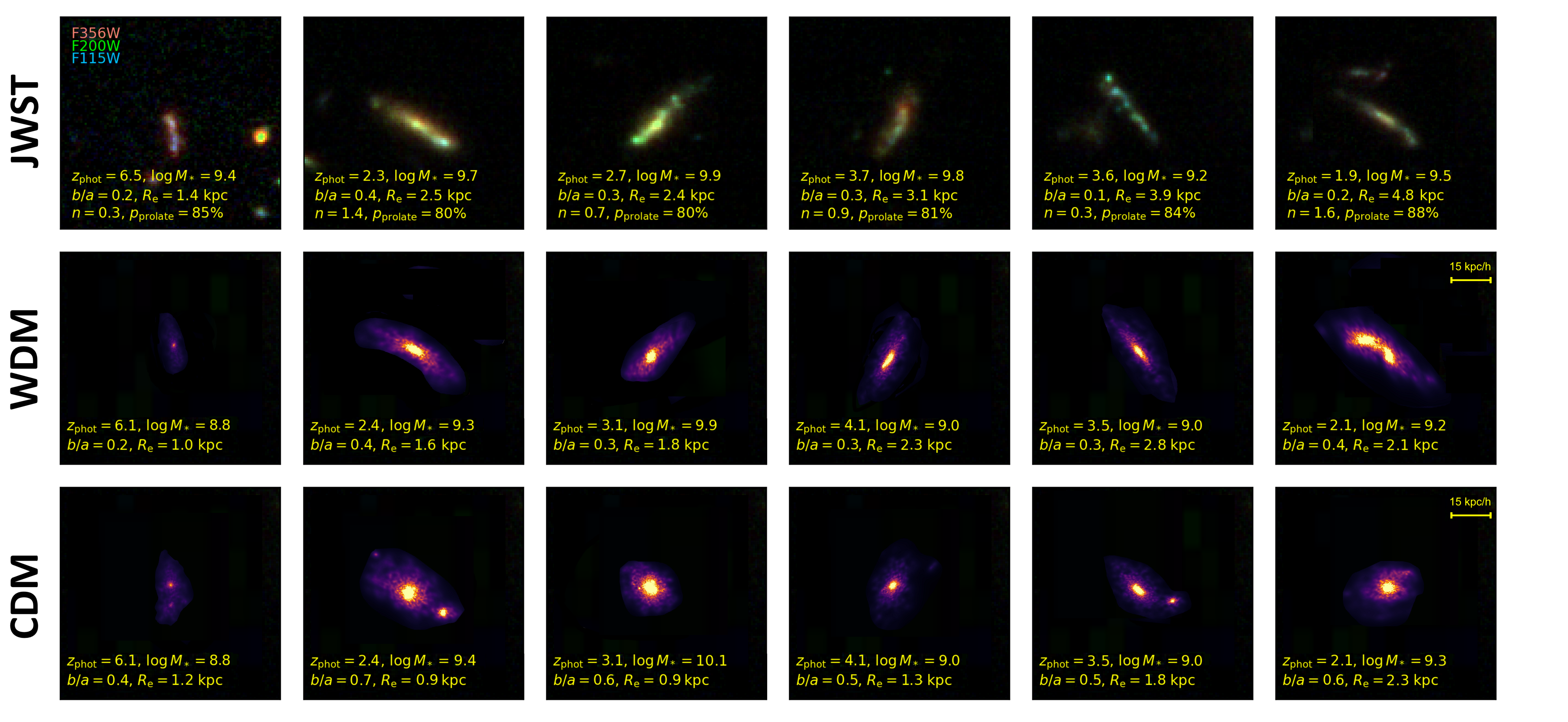}
	\caption{{\bf Illustrative comparison of high-z galaxies detected by JWST and simulated galaxies}. The first row shows representative $3^{\prime\prime}\times3^{\prime\prime}$, observed examples of the prolate class reproduced from Ref\cite{Pandya:2024}, SE++ catalog. The next two rows show the predicted stellar appearance of simulated galaxies for the two different classes of dark matter from our hydro-simulations. For each column the simulated galaxy locates at the same position in the simulation volume for a fair comparison of all two DM models with matching initial conditions. In each model, we identify a simulated galaxy with similar stellar mass and project it to best match the observed data. Note also that we compare the same span of redshift. We have matched the observed orientation but maintained the size and stellar density range equally between the simulated galaxies. }\label{bananas}
    
\end{figure}

\begin{figure}[H]
    \centering
 \includegraphics[width=180mm,height=16.5cm]{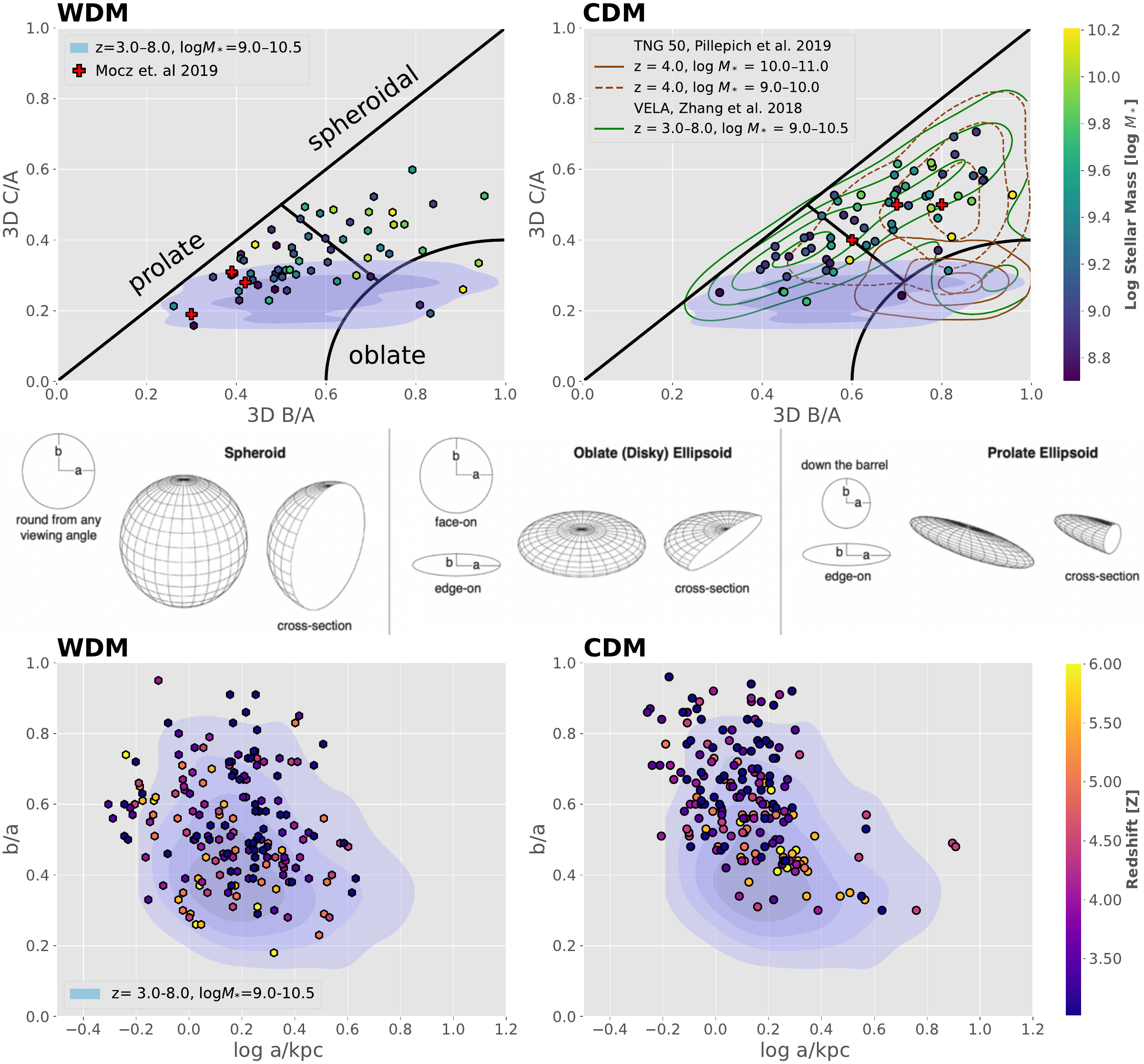}
 
    \caption{ {\bf 2D \& 3D stellar morphology of simulated galaxies compared with observations $3<z<8$}. All time frames between $z = 3$ and $z = 6$ are shown, where the color scale indicates stellar mass $M_{*}$ (top) and redshift (bottom), with overlapping points offset slightly for clarity. Note that the redshifts ($z$) and stellar masses ($M_{*}$) of the simulated halos match those of the observations. The left panels present WDM results, while the right panels show galaxy shapes predicted for CDM. {\bf Top row:}  The black boundaries define 3D ellipsoids as oblate, spheroidal, or prolate, following Refs\cite{Wel:2014,Zhang:2019}. Colour scales indicate the stellar number weighted age of the stars in each simulated galaxy, determined at each time frame of the simulations. The shaded areas represent the distribution of JWST based measurements from Ref\cite{Pandya:2024}. Red data points represent independently calculated ellipticities of the dark matter halos by Ref\cite{Mocz:2020}. Each point represents a galaxy, although in some cases the same galaxy is shown at multiple redshifts. In total, the sample includes as few as two distinct galaxies at $z = 6$ and up to 18 at $z = 3$. The solid and dashed brown contours represent the predicted distribution for CDM from TNG50 \cite{Pillepich:2019} and the green ones from VELA \cite{Zhang:2019}.  {\bf Middle row:} Helpful representation of the classification of ellipsoid shapes, as depicted by Ref\cite{Pandya:2024}. {\bf Bottom row:} Projected semi-axis ratio, $b/a$, versus projected semi-major axis $a$, for comparison with the observations represented by shaded areas for z $>$3 \cite{Pandya:2024}. This shows the larger spread towards smaller b/a with WDM, similar to the data when the simulated galaxies are young and lower mass, as indicated by the color bar. The number of data points here is several times larger than in the upper panel because we show three orthogonal projections for each galaxy.}

    \label{Fig:compnew}
\end{figure}

Figure \ref{bananas} shows an illustrative comparison between typical galaxies from Ref\cite{Pandya:2024} (top row) and the visual appearance of our simulated galaxies (middle row: WDM; bottom row: CDM). The observed galaxies are noticeably elongated and we readily find similar-looking simulated galaxies for WDM, whereas for CDM, the stellar morphologies of the corresponding simulated galaxies are noticeably rounder with additional merging sub-halos that are less like the observed examples. In general, smooth and elongated galaxies are not common in our CDM simulations. Previous CDM simulations have not presented a consensus in this respect, with the VELA simulations predicting a significant number of prolate galaxies at high redshift \cite{Tomassetti:2016}, whereas for the hydro-CDM TNG50 simulation, Figure 9 of Ref\cite{Pillepich:2019} predicts an appreciable fraction of prolate-looking galaxies for low-mass systems at z$>$4; oblate galaxies dominate their simulation, in good quantitative agreement with our CDM predictions (see top right panel of Figure \ref{Fig:compnew}). 

We note that the simulated galaxies span the same redshift range as the observational data\cite{Pandya:2024}, and cover the same stellar mass range, with 90\% of our simulated galaxies in the range $10^9 < M_{*} < 5 \times 10^{10}\,M_{\odot}$. We also find that the predicted ellipticity is not a strong function of redshift, but is primarily influenced by the presence of long filaments, as described fully below.

We quantify the 2D (projected) and 3D ellipticities of the stellar distributions within each predicted galaxy for each class of dark matter following the methods of Ref\cite{Pandya:2024}, using the Minimum Volume Enclosing Ellipsoid (MVEE) method in Python. In 3D, we estimate stellar number-weighted axis ratios using the principal axes obtained from the ellipsoid fit, while in 2D we apply the same procedure to the projected stellar positions to derive the corresponding ellipse parameters (see Methods). The first row of Figure~\ref{Fig:compnew} compares the 3D axis ratio distributions of our simulated galaxies (points) with the observationally inferred distributions (shaded areas) from Ref\cite{Pandya:2024} for $z>3$, with low-, medium-, and high-mass galaxies ($10^9 - 5\times 10^{10}\,M_{\odot}$) colour-indicated. The black boundaries classify 3D ellipsoids into prolate, oblate, and spheroidal shapes following Ref\cite{Wel:2014} and Ref\cite{Zhang:2019}, as adopted by Ref\cite{Pandya:2024}. For WDM, most simulated galaxies fall in the prolate class ($b/a \approx 0.2$--$0.4$), especially at high redshifts, with this tendency persisting up to $z=3$ and matching well the observed projected distributions. In contrast, CDM galaxies are predominantly spheroidal across all redshifts and stellar masses, with only a few prolate systems. Note, the bulk of observed CEERS galaxies lie well above the PSF limit, as shown in Figure 4 of Ref\cite{Pandya:2024}, with $2r_e > $ PSF-FWHM, which is also similar to our WDM predictions, whereas a large fraction of our CDM galaxies are small, requiring PSF convolution and pixel binning for a more accurate comparison.

\begin{figure}[H]
    \centering
 \includegraphics[width=180mm,height=17.5cm]{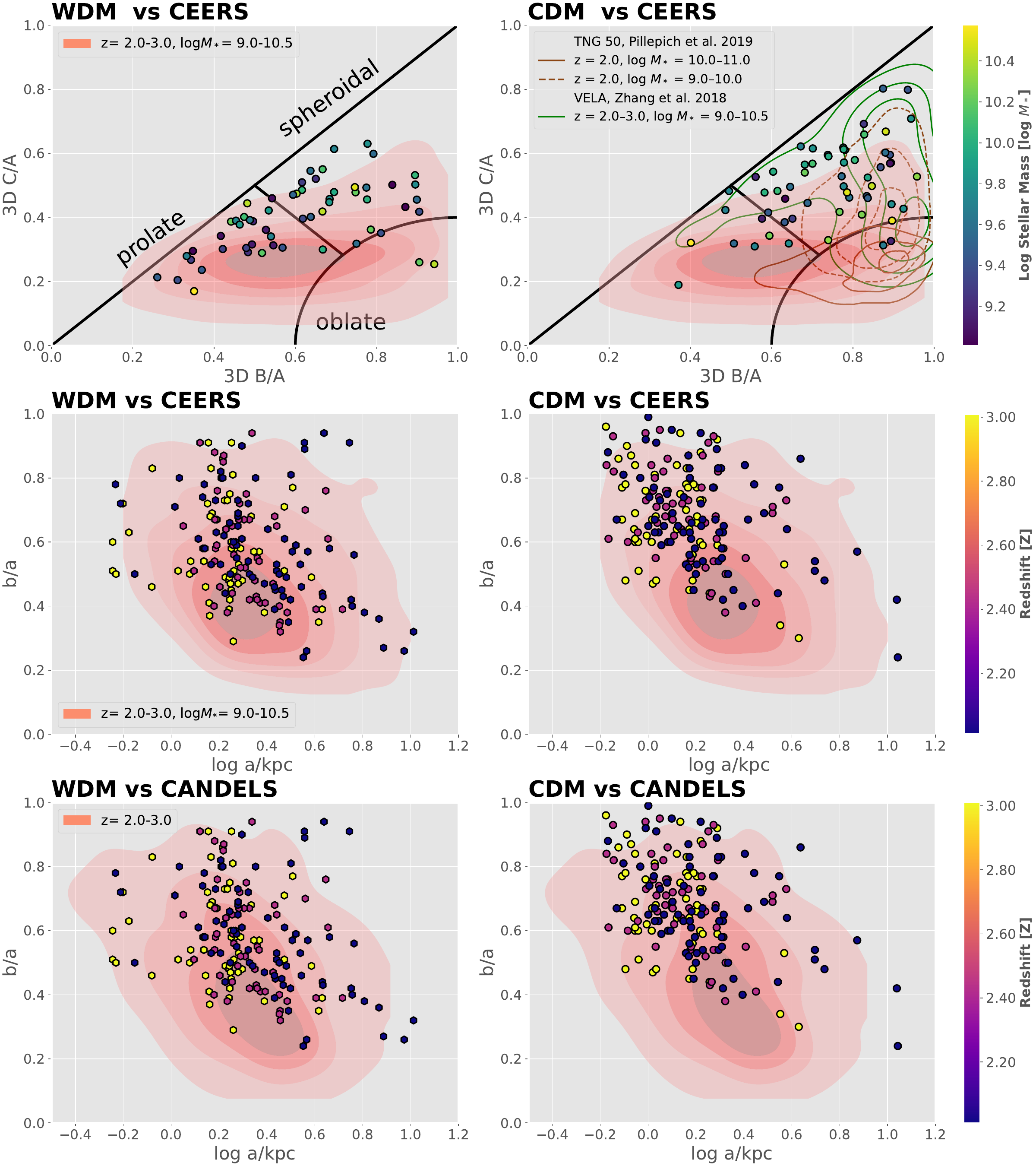}

-    \caption{ {\bf 2D \& 3D Stellar morphology of simulated galaxies at $2<z<3$}. The points represent simulated galaxies of WDM and CDM for comparison with contours representing the observed CEERS and CANDELS surveys, where the redshift ($z$) and stellar mass ($M_{*}$) range of the simulations match the observed ranges. \textbf{Top row:} The black boundaries classify 3D ellipsoids as oblate, spheroidal, or prolate, following the criteria of Refs\cite{Wel:2014, Zhang:2019}. The red shaded area represents the distribution of measurements based on JWST data from Refs\cite{Pandya:2024} for CEERS galaxies with $z = 2-3$. The solid and dashed brown contours represent the predicted distribution for CDM from TNG50 \cite{Pillepich:2019} and the green ones from VELA \cite{Zhang:2019}. Each point in the figure represents a galaxy, although in some cases the same galaxy is shown at multiple redshifts. In total, the sample includes 18 distinct galaxies at $z = 3$ and up to 25 at $z = 2$. The solid and dashed black contours represent the CDM hydro-simulation predictions of TNG50 \cite{Pillepich:2019}. \textbf{Middle row:} Projected ($b/a$) vs. semi-major axis ($a$) for comparison with the red shaded area for $z = 2-3$ of the CEERS survey \cite{Pandya:2024}. The colored data points correspond to the three orthogonal projections of each simulated galaxy. The sample includes projections of 18 galaxies at $z = 3$ and up to 25 halos at $z = 2$.
\textbf{Bottom row:} Similar to the middle row, where the red contours now represent the CANDELS survey \cite{VDW:2014} galaxies with $z = 2-3$. } 


    \label{Fig:compnew2}
\end{figure}

In Figure \ref{Fig:compnew2}, we extend the analysis to lower redshifts $2 < z < 3$ so we can compare with the independent CANDELS/HST survey\cite{Grogin,VDW:2014}. Again, the WDM predictions show better agreement with the data than CDM, with semi-axis ratio values more accurately reproducing the observed peak at $b/a = 0.4$, whereas CDM tends to produce galaxies that are smaller than observed. 

\begin{figure}[H]
    \centering
 \includegraphics[width=180mm,height=6.5cm]{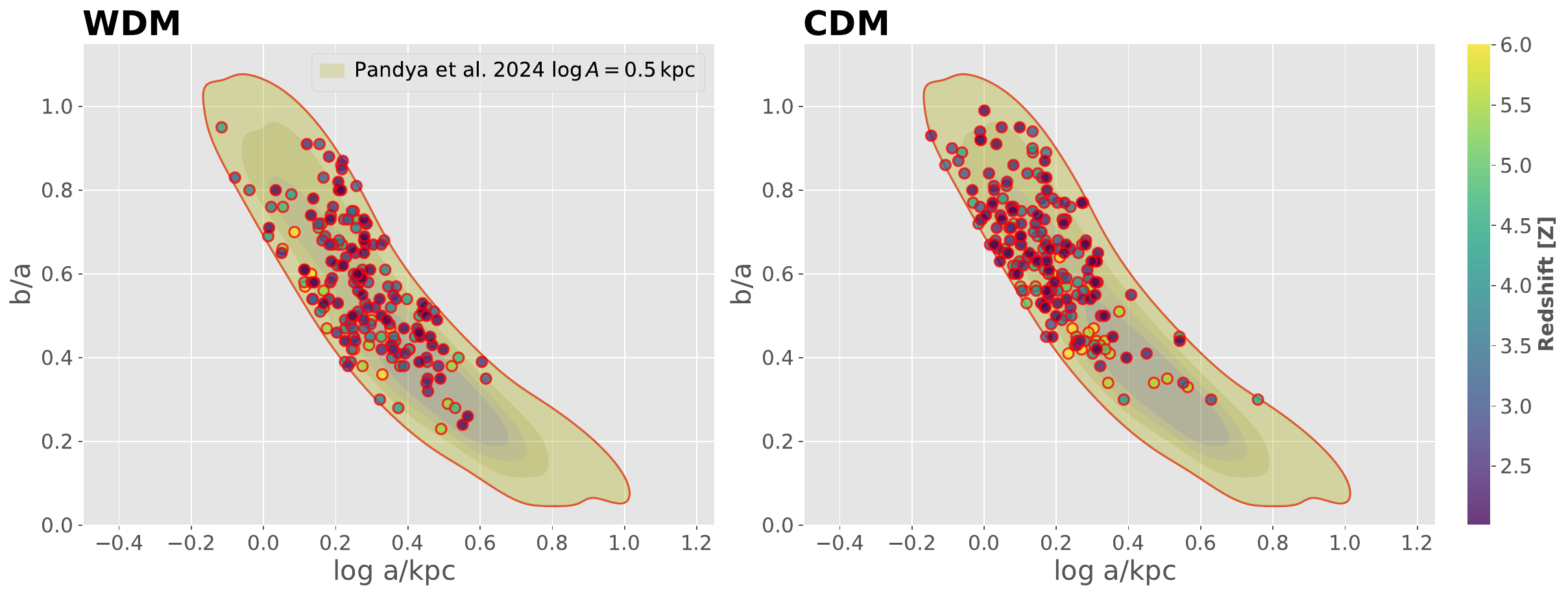}
    \caption{ {\bf Projected semi-axis ratio, $b/a$, vs. projected semi-major axis shown for the stellar distribution of each galaxy.} \textbf{Left panel:} results for the WDM simulation, including all three orthogonal projections for $2 < z < 6$ (as presented in Figure~\ref{Fig:compnew}). We show galaxies enclosed within the “banana”-shaped region (yellow-shaded area and red contour) predicted for a prolate population of triaxial ellipsoids in projection, assuming a fixed 3D major axis of $\log A/{\rm kpc} = 0.5$ and a Gaussian spread of 0.03\,dex, following Ref.~\cite{Pandya:2024}. The color of each point indicates the redshift, as in Figure~\ref{Fig:compnew}. \textbf{Right panel:} same as in the left panel, but for galaxies in the CDM simulation. All other conventions are identical.
}

    \label{Fig:banana1}
\end{figure}

To support this interpretation, we calculate the log-likelihood differences between the two models across all figure rows in Figures \ref{Fig:compnew}, \ref{Fig:compnew2} and \ref{Fig:banana1}. The methodology, including the treatment of correlations between projections of the same halos in the 2D case, is described in the Methods section, and the corresponding values are listed in Table~\ref{tab:1}. For each row, we computed the total log-likelihood ($\ln \mathcal{L}{\mathrm{tot}}$) of WDM and CDM with respect to the corresponding observational dataset (left and right panels, respectively), summing over all galaxies. The difference $\Delta\ln \mathcal{L}{\mathrm{tot}} = \ln \mathcal{L}{\mathrm{tot}}^{\mathrm{WDM}} - \ln \mathcal{L}{\mathrm{tot}}^{\mathrm{CDM}}$ was then calculated separately for each row (e.g. CEERS–2D, CEERS–3D, CANDELS–2D in Figure \ref{Fig:compnew2}). In nearly all comparisons, the \emph{total} log-likelihood values are higher for WDM than for CDM, indicating very strong statistical evidence ($\Delta \ln \mathcal{L}{\mathrm{tot}} > 10$). The exception is the 2D comparison shown in the bottom panel of Figure~\ref{Fig:compnew}, where the difference ($\Delta \ln \mathcal{L}_{\mathrm{tot}} = 6.24$) still corresponds to strong statistical evidence in favour of WDM. According to standard interpretations of likelihood ratios, differences above $\Delta \ln \mathcal{L}{\mathrm{tot}} \approx 5$ indicate strong to decisive evidence, while values exceeding 10 correspond to overwhelming evidence for one model over the other. The fact that WDM consistently outperforms CDM across all comparisons — summarized in Figure~\ref{Fig:curve}, further illustrated by the simulations in the Supplementary Material, visually noticeable in Supplementary Figure 1 (with many more prolate galaxies such as those shown in the first row of Figure~\ref{bananas}), and statistically supported by Supplementary Figures 2 and 3 — combined with the large differences found in most cases, suggests a systematic and statistically significant preference for WDM.


We note that the level of agreement between WDM and CDM and the data is slightly weaker for CANDELS/HST Figure~\ref{Fig:compnew2} than for CEERS/JWST in Figure~\ref{Fig:compnew}. This may be due to the limited resolution of the baryonic component, which is constrained to $512^3$ particles per snapshot, and motivates a consistency check simulation performed at higher resolution but with a smaller box at low redshift and lower stellar masses. This is presented in the Supplementary Material, and produces a consistent trend with the simulations shown here. 

Comparison of our CDM simulations with the hydro simulations of TNG50 and VELA shows good agreement with the predicted 3D stellar ellipticities, regardless of stellar mass or redshift, as shown in Figures \ref{Fig:compnew} and \ref{Fig:compnew2}. Our simulations also follow the observed trend in the fraction of prolate galaxies, which is lower for CDM at lower redshift. The VELA hydro-CDM simulations \cite{Ceverino:2015} are closer to the data at $z=2$ (see Fig. 17 of Ref\cite{Zhang:2019}), although their average value of $\langle b/a \rangle = 0.59$ is more consistent with other CDM predictions than it is with CANDELS and CEERS, which show lower mean values of $\langle b/a \rangle \approx 0.48$ at similar stellar masses and redshifts (see Fig. 16 of Ref\cite{Zhang:2019}). However, VELA also predicts a minority population of prolate galaxies, especially at $z>3$, as seen in the top panel of Figure \ref{Fig:compnew}, matching the range suggested by the data similarly to WDM. This indicates that hydro-CDM simulations are able to reproduce a notable number of prolate galaxies, as required by the observations. At the same time, Figure~\ref{Fig:compnew} shows that VELA and TNG diverge in their predictions at higher stellar masses for $z>3$, whereas our CDM simulations remain in closer agreement with VELA over the entire redshift range (see also Figure~\ref{Fig:compnew2}). This highlights that there are at least two CDM-based models with size--shape distributions that do not agree with each other, suggesting that the hydrodynamical method (AMR in VELA and AREPO for TNG50) and/or the adopted subgrid physics may significantly impact the predictions.

Figure \ref{Fig:banana1} shows the ratio of semi-minor to semi-major axes, b/a, vs. semi-major axis, a, for the stellar content of galaxies in the simulations. The distribution of b/a for WDM can be seen to extend to lower values than for CDM, where most galaxies are rounder in terms of the projected stellar ellipticity. In Figure~\ref{Fig:curve}, we examine the dependence of $b/a$ across the full redshift ranges of our three simulations (including $\psi$DM in the right panel). We use the mean $b/a$ values estimated for CEERS~\cite{Kartaltepe:2023, Pandya:2024}. The left panel shows results from the main 10~Mpc/$h$ simulation box as well as from the 5~Mpc/$h$ box that is presented in more detail in Section 1 of the Supplementary Material; these provide an overview of the full range of halo and stellar masses. The right panel shows results from the smaller 1.7~Mpc/$h$ boxes from Ref~\cite{Mocz:2020}, though this box size is less suitable for direct comparison with observations due to the more limited overlap in stellar mass and redshift, it allows us to present $\psi$DM simulations. In both panels, the halo and stellar mass ranges are included in the figure caption. 

Figure~\ref{Fig:curve} shows that the trend predicted for CDM is consistently above that of WDM/$\psi$DM, with the observational data appearing to favor the lower values of b/a of WDM/$\psi$DM. This agreement is also reflected in the universally lower $\chi^2$ values shown in the panel legends. The mass range is largest for CDM, where galaxies form earliest and smallest for $\psi$DM, which is numerically limited to $z > 5.5$. These discrepancies between the different dark matter models, where WDM/$\psi$DM are favored by the data, can also be observed statistically in Supplementary Figure 4 and visually in Supplementary Figure 5.

Improving these comparisons will require better $\psi$DM and WDM simulations that explore a broader range of particle masses. In addition, deeper imaging from JWST at higher redshifts will be crucial and will require accurate PSF corrections. The two additional simulations are less suitable than the 10~Mpc/$h$ run for direct comparison with the observational data of Ref~\cite{Pandya:2024}, owing to more limited overlap in stellar mass and redshift. The 1.7~Mpc/$h$ box simulation of Ref~\cite{Mocz:2020}, performed several years before the advent of JWST and thus representing genuine priors, highlights the close agreement between WDM and $\psi$DM and their shared contrast with CDM. Overall, these simulations confirm that the tendency toward greater elongation is a robust feature of the smooth filaments in WDM and $\psi$DM, largely independent of galaxy mass, and consistent with the trends identified in the main simulations.




\section{Discussion and Conclusions} \label{sec:cite}

We have made a straightforward comparison of {\it stellar} morphology predicted by hydrodynamical simulations of three contending DM models,  CDM, WDM \& $\psi$DM, finding that WDM and $\psi$DM are similar in terms of predicting mainly prolate stellar morphology during the first $0.5$ Gyrs after formation, whereas for CDM the stellar morphology is typically spheroidal or oblate. We conclude that the observed predominance of prolate-shaped galaxies recently derived from JWST images of galaxies at $z>3$ by Ref\cite{Pandya:2024} supports equally WDM or $\psi$DM but is in tension with CDM (although we note that some CDM hydrodynamical simulations, such as VELA, can produce a notable proportion of prolate galaxies for high redshifts), and this is also supported by comparison with the CANDELS survey at $z>2$. Similar tension with CDM at low redshift has also been recently claimed, with an excess of elongated low-mass galaxies in the GAMA survey~\cite{GAMA} relative to the FIREbox hydro-simulation of CDM~\cite{FIREBOX}. Such elongated, low-mass galaxy analogues are also found to be typically prolate dynamically~\cite{local,Valle:2023} rather than rotating edge-on disks.

The prolate shapes predicted by our WDM and $\psi$DM simulations form from long, smooth filaments that lack small-scale power and thus substructure is absent along the filaments. These filaments of WDM and $\psi$DM form first in our simulations, ahead of galaxy formation, in an early period uninterrupted by merging, until approximately $z\simeq 4$ below which the first galaxy mergers become common for WDM and $\psi$DM, as emphasized by Ref\cite{Schive:2016}. Direct estimates are now available for merging from the rate of close pairs of galaxies detected by JWST \cite{Duan:2024,Pusk}, indicating that merging affects a significant minority ($\simeq 10\% $) of galaxies by $z \simeq 6$. The first constraints at higher redshift may indicate a turnover\cite{Pusk} that can be compared with the simulations reported here, to help distinguish DM models. For CDM, several luminous subhalos are typically predicted to be orbiting young galaxies at high-z, as merging proceeds within the first Gyr. Such subhalos are potentially resolvable with upcoming laser-based instruments \cite{Windhorst:2024}. For this purpose it will be important to explore a range of feedback models affecting subhalo detectability\cite{heating1,heating2}, which in our current simulations is simply implemented at the sub-grid level to approximate SNII gas outflows \cite{Vogelsberger:2013}. 
 
 Our work motivates a dedicated search for an early filament era, predicted prior to galaxy formation for both $\psi$DM and WDM, that may be traced by ionized gas, HI and stars\cite{Gao:2007,Mocz:2019,F2,F3}. Direct detection of such filaments will be challenging with low levels of star formation, but is perhaps hinted in the recently observed pair of elongated, very young galaxies at z$\sim$8.3 in Fig.~2 of Ref\cite{Ma:2024}). An elongated object of around 20 kpc has also been detected recently in deep JWST/NIRSpec observations at z$\sim$6.2, illuminated centrally by a young quasar \cite{Loiacono:2024}, validating the filament origin proposal of Ref\cite{Pandya:2019} made prior to JWST. Filament length can also provide information on the DM particle mass and in the simulations here, particle masses 0.91 KeV were adopted, equivalent to $0.8\times 10^{-22}$ eV and $2.5\times 10^{-22}$ eV for WDM and $\psi$DM respectively, corresponding to a comoving filament length of $\simeq 150$ kpc at $z\simeq 10$, or about 0.5 arcminutes, well suited to the field of JWST. Such a measurement can complement the core-halo analysis reported for local group dwarfs in the context of $\psi$DM, where a soliton scale of about 0.5 kpc has been identified\cite{Pozo:2020,Pozo:20232} and can be compared to the relatively thick band (4 kpc) of abundant microlensed stars in JWST monitoring of the ``Dragon Arc" \cite{Fudamoto,Yan} where consistency has been claimed with the corrugated critical curves of $\psi$DM for a boson mass of $\simeq 2\times10^{-22}$ eV \cite{Broadhurst:2024}.

In addition to Ref\cite{Pandya:2024}, which has principally motivated our analysis, there is independent JWST evidence for a high proportion of elongated galaxies, initially classed as edge-on disks or peculiar\cite{Ferreira:2022}, and also an enhanced evolving abundance of elongated galaxies recognised at high-z in the EPOCHS survey\cite{Excess} and a class of "ultra red flattened" prolate galaxies recognised in the JADES survey\cite{Gibson:2024}. Hence, we conclude that galaxy formation originating from smooth filaments provides a natural explanation for the observed predominance of elongated galaxies at young ages, motivating the consideration of a wider range of physical possibilities for dark matter that are now becoming testable with JWST.

\section{Methods} \label{sec:cite}

\subsection{Numerical methodology: Simulation}

We adopt the magneto-hydrodynamics solver \texttt{AREPO}\cite{Springel:2010} as the main numerical tool for cosmological simulations. The code solves hydrodynamical equations for gas physics on an adaptive moving mesh. For the CDM and WDM model, dark matter are treated as collisionless particles with gravitational forces calculated via particle-mesh and tree algorithms. For the $\psi$DM model, since dark matter is wave-like with its dynamics governed by the Schrödinger-Poisson equations~\cite{},
\begin{equation}
    i\hbar \dfrac{\partial\psi}{\partial t} = \dfrac{\hbar^2}{2ma^2}\nabla^2\psi + \dfrac{m}{a}\Phi\psi; \qquad \nabla^2\Phi = 4\pi G\left( |\psi|^2 - \langle |\psi|^2 \rangle \right),
\end{equation}
simulations with this type of dark matter need a separate treatment for gravity. For that reason, we modify \texttt{AREPO} to include the pseudospectral solver for $\psi$DM in cosmology, originally developed by Ref\cite{Mocz:2017}, which yields results converging at second order in time and exponentially in space. However, as the method strictly applies to uniform grids, it is limited in finest details that can be resolved, even with the highest archivable resolution with our computing resources.

Initial conditions given in terms of positions and velocities of these particles are generated with \texttt{N-GenIC}\cite{Springel:2015} from a given matter power spectrum that are computed by \texttt{CAMB}\cite{Lewis:2011} for CDM and by \texttt{axionCAMB}\cite{Hlozek:2015} for WDM/$\psi$DM. The hydrodynamical setup is identical across simulations of different dark matter species in the same box size, with the initial perturbations differing only in their corresponding power spectrum shape. In CDM, no cutoff is applied, while WDM and $\psi$DM assume exponentially suppressed initial power spectrum. However, the WDM simulation does not account for the initial velocity dispersion of WDM particles, differing from classical WDM models. Instead, the WDM case is designed to be ``$\psi$DM minus wave effects'', a widely used approximation in cosmological studies.

The baryonic feedback we incorporate is modeled at the sub-grid level within the \texttt{AREPO} code, exactly as implemented for the Illustris-TNG project \cite{Springel:2018} with an appropriate scaling of some parameters regarding softening scales and AGN feedback to account for different box sizes and particle numbers.
The effect of stellar feedback in relation to stellar sphericity has been specifically investigated by Ref\cite{Tomassetti:2016} in earlier hydro-CDM simulations and found to possibly delay transition from an early prolate phase to later oblate disk shape. The formation of very early filaments formed in $\psi$DM/WDM have been attributed by Ref\cite{Mocz:2020} to the dominance of the central relatively deep linear potential over feedback effects, at least at early times when the filaments are initially visible. The effect of varying the strength and nature of feedback may well be expected to have evolutionary consequences and can be examined in greater detail with dedicated new simulations in relation to the formation time of prolate structure and its evolution towards more spheroidal and oblate morphology\cite{feedback}.

Here we compare three different simulations, one in the main body and the other two in the supplementary material, each performed with different box sizes of 10, 5, and 1.7\,Mpc/$h$, and exploring three dark matter candidates: Cold Dark Matter (CDM), Wave/Fuzzy Dark Matter ($\psi$DM), and Warm Dark Matter (WDM, often used as a proxy for $\psi$DM). The 10\,Mpc/$h$ and 5\,Mpc/$h$ simulations were specifically carried out for this work, while the 1.7\,Mpc/$h$ simulation was presented in the earlier study by Ref\cite{Mocz:2020}
. The simulation with the largest volume (10 Mpc/h) is presented in the main body of the paper, while the smaller-box simulations are included in the supplementary material, 
exploring lower galaxy masses with higher resolution. The main-body simulation for WDM and CDM allows for a comparison at the higher stellar masses of the observational data of the CEERS and CANDELS surveys. The “WDM” simulations analyzed here were performed using a $\psi$DM initial power spectrum, without including the dynamical quantum potential. This approximation reproduces the small-scale cutoff expected for a thermal relic “WDM” particle of mass 0.91 keV, corresponding to the boson mass of $0.8\times10^{-22}$ eV used in the $\psi$DM simulations of Ref\cite{Pozo:2021}. The motivation for CDM and WDM is of course long-standing, but diminishing with increasingly stringent laboratory absence of any heavy particle DM contender or direct underground detection \cite{Aprile:2022}, including WIMP's at the TeV range, or new keV mass scale particles needed for WDM. However, WDM remains physically viable in terms of hypothetical right-handed ``sterile" neutrinos as these naturally evade Standard Model interactions provided they are light enough that early relativistic free streaming results in sizeable DM cores and significantly suppresses lower mass galaxies.

\subsubsection{Setup of the 10 Mpc/h Cosmological Simulation}

The cosmological simulation presented in this main body of the paper was performed for a volume with a size of \( L_{\text{box}} = 10 \, h^{-1} \, \text{Mpc} \), assuming WDM particle mass of 0.91 keV equivalent to a boson mass of \( m = 0.8 \times 10^{-22} \, \text{eV} \). This choice introduces a cutoff in the initial power spectrum at \( L_{\text{cutoff}} \approx 2.5 \, h^{-1} \, \text{Mpc} \) due to the uncertainty principle. The simulation evolves from redshift \( z = 127 \) (corresponding to a universe age of \( 10^7 \, \text{years} \)) to redshift \( z = 2.01 \) (universe age \( 10^9 \, \text{years} \)), yielding between 2 galaxies at \( z = 6 \) and 30 at \( z = 2 \).
 The dark matter and baryonic resolution is \( 512^3 \) particles, corresponding to a baryonic mass resolution of approximately \( 8.6 \times 10^4 \, M_\odot \) and \( 1.6 \times 10^4 \, M_\odot \) specifically for stars . The simulation adopts cosmological parameters measured by the \textit{Planck} satellite \cite{Planck:2016}, with \( \sigma_8 \) boosted from 0.8 to 1.0 in comparison to the value from Ref\cite{Naoz:2012}. This enhancement compensates for the limited cosmological volume for the required halos stellar masses, probed and increases the initial fluctuations \cite{Naoz:2012}. The warm dark matter simulations are compared to cold dark matter simulations, both of which are run with a resolution of \( 512^3 \) dark matter particles.

\subsubsection{Setup of the 5 Mpc/h Cosmological Simulation}

This section describes larger hydro-simulations feasible for WDM and CDM (but not $\psi$DM),  that extend to lower stellar masses, ($10^7-10^9 M_{\odot}$) to allow an additional comparison with the CANDELS and CEERS surveys at $z>2$. These extra simulations have a smaller box size of $5.1 \, h^{-1} \, \mathrm{Mpc}$, with a $512^3$ resolution used for dark matter and baryon gas particles, to give a mass resolution of $1.14 \times 10^{4} \, M_{\odot}$. The increased resolution enables more detailed galaxy morphologies and allows for a convergence test with our larger-volume simulation. The simulation evolves from redshift $z = 127$ to $z = 1.01$, adopting cosmological parameters measured by the \textit{Planck} satellite\cite{Planck:2016}, with $\sigma_8 = 0.8$, in agreement with the value recommended by Ref\cite{Naoz:2012}.
 The revised setup yields halos with virial masses of $10^8 - 10^{11} M_{\odot}$ but with lower stellar masses in comparison to the data from Ref\cite{Pandya:2024}, $10^7 - 5\times 10^{9}\,M_{\odot}$, as most of the resulting halos in this setup do not reach $10^9\,M_{\odot}$, although those shown in the redshift range $2 < z < 3$ do exceed $10^8\,M_{\odot}$.
\subsubsection{Setup of the 1.7 Mpc/h Cosmological Simulation}

Here a cosmological simulation for $\psi$DM was performed in addition to WDM and CDM~\cite{Mocz:2019} in a periodic volume with a size of \( L_{\text{box}} = 1.7 \, h^{-1} \, \text{Mpc} \), assuming a boson mass of \( m = 2.5 \times 10^{-22} \, \text{eV} \). This choice introduces a cutoff in the initial power spectrum at \( L_{\text{cutoff}} \approx 1.4 \, h^{-1} \, \text{Mpc} \) due to the uncertainty principle. The simulation evolves from redshift \( z = 127 \) (corresponding to a universe age of \( 10^7 \, \text{years} \)), to redshift \( z = 5.5 \) (universe age \( 10^9 \, \text{years} \)). The final redshift is constrained by resolution requirements to ensure fully converged results. The $\psi$DM dark matter spectral resolution is \( 1024^3 \) , while the baryonic resolution is \( 512^3 \) particles, corresponding to a mass resolution of approximately \( 2.64 \times 10^3 \, M_\odot \). The simulation adopts cosmological parameters measured by the \textit{Planck} satellite \cite{Planck:2016}, with \( \sigma_8 \) boosted from 0.8 to 1.4 in comparison to the value from Ref\cite{Naoz:2012}. This enhancement compensates for the limited cosmological volume probed and increases the initial fluctuations \cite{Naoz:2012}. The $\psi$DM simulation is compared with CDM and WDM simulations, both of which are run with a resolution of \( 512^3 \) dark matter particles. To relate particle masses in the $\psi$DM and WDM cosmologies, the cutoff scale is matched, yielding a WDM particle mass of approximately \( m_{\text{WDM}} \sim 1.4 \, \text{keV} \). This setup yields halos that closely with virial masses of $5\times 10^7 - 5\times 10^{10}\,M_{\odot}$ but with lower stellar masses in comparison to the data from Ref\cite{Pandya:2024}, $5\times 10^6 - 5\times 10^{8}\,M_\odot$.

\subsection{Extracting morphology of the galaxies}

To characterize the projected morphology of the galaxies, we model their stellar distributions as ellipses on the plane of the sky. In this framework, each galaxy is described by its semi-major (a) and semi-minor (b) axes, as well as the coefficients (A, B) that define the orientation and shape of the ellipse in Cartesian coordinates. These parameters are derived from the second moments of the projected stellar distribution, providing a robust estimate of the overall shape of each galaxy.

\subsection*{Minimum Volume Enclosing Ellipsoid (MVEE) and Elliptical Shape Estimation}

To robustly characterize the projected morphology of each galaxy, we model the stellar distribution as an ellipse and estimate its shape parameters using the \textit{Minimum Volume Enclosing Ellipsoid} (MVEE). This method computes the unique ellipsoid of smallest volume that contains all (or a filtered subset of) the projected stellar positions. MVEE-based fitting is widely used in applications ranging from multivariate data analysis to computational geometry, as it provides a compact and noise-resistant representation of the spatial extent of a point distribution \cite{Vandenberghe:1998}.

We adopt a publicly available Python implementation of the MVEE algorithm, based on Khachiyan’s iterative method, which converges efficiently for moderate-size point sets. Given a 2D array of projected stellar coordinates \((x_i, y_i)\), the algorithm returns the center of the ellipsoid \((x_0, y_0)\), the lengths of the semi-major and semi-minor axes \((a, b)\), and the orientation angle \(\theta\) (or equivalently, the ellipse coefficients \(A\) and \(B\) in quadratic form). For visualization and verification, we make use of the \texttt{matplotlib.patches.Ellipse} class.

\paragraph{Preprocessing: Density-Based Filtering.}

To ensure that the computed ellipse reflects the \textit{bulk} of the stellar distribution and is not biased by tidal features, stellar streams, or isolated outliers, we apply a density-based filtering step. Specifically, we discard the 10\% of stars located in the least dense regions of the projected stellar map. Local stellar density is estimated either via a fixed number of nearest neighbors or kernel density estimation (KDE). This step effectively confines the MVEE fit to the dense central region of the galaxy, improving robustness across diverse morphologies and mass ranges.

\paragraph{Stellar Axis Ratios and Effective Radius.}

Once the projected ellipse is determined, we extract the semi-major axis \(a\), semi-minor axis \(b\), and compute the axis ratio \(q = b/a\). These geometric parameters provide a model-independent estimate of the projected flattening of each galaxy.

We also compute star number-weighted effective radii based on the derived elliptical shape. Since the area of an ellipse is given by \(\pi a b\), and assuming a circularized radius \(r_{\rm eff}\) such that \(\pi r_{\rm eff}^2 = \pi a b\), we define:

\[
r_{\rm eff}^2 = a_{\rm eff} \times b_{\rm eff}
\]

Given a measured axis ratio \(q = b/a\), we can express:

\[
b_{\rm eff} = q \times a_{\rm eff} \quad \Rightarrow \quad a_{\rm eff} = \frac{r_{\rm eff}}{\sqrt{q}}
\]

This allows us to recover the star-weighted semi-major axis \(a_{\rm eff}\), which we use as a size proxy for morphological comparisons and radial profile extractions.

\subsection*{3D Ellipsoidal Fit: Extraction of Principal Axes}

In three dimensions, the Minimum Volume Enclosing Ellipsoid (MVEE) provides a compact and orientation-aware fit to the outer stellar distribution. The algorithm returns a symmetric positive-definite matrix \( A \in \mathbb{R}^{3 \times 3} \) and a center \( \mathbf{c} \in \mathbb{R}^3 \) such that the ellipsoid is defined by:

\[
(\mathbf{x} - \mathbf{c})^T A (\mathbf{x} - \mathbf{c}) = 1
\]

This quadratic form represents a rotated ellipsoid centered at \( \mathbf{c} \), with shape and orientation encoded in the matrix \( A \). To extract the physical semiaxes, we diagonalize \( A \) using a singular value decomposition (SVD) or eigendecomposition:

\[
A = V^T \Lambda V
\]

where \( V \) is an orthonormal matrix whose columns are the principal directions (eigenvectors), and \( \Lambda = \mathrm{diag}(\lambda_1, \lambda_2, \lambda_3) \) contains the eigenvalues of \( A \).

Since the standard form of a 3D ellipsoid aligned with the coordinate axes is:

\[
\left( \frac{x}{r_1} \right)^2 + \left( \frac{y}{r_2} \right)^2 + \left( \frac{z}{r_3} \right)^2 = 1
\]

we identify the semiaxes \( r_i \) with:

\[
r_i = \frac{1}{\sqrt{\lambda_i}} \quad \text{for } i = 1, 2, 3
\]

The resulting values \( r_1, r_2, r_3 \) correspond to the lengths of the principal axes of the ellipsoid. We denote these by \( A, B, C \), sorted in descending order such that \( A \geq B \geq C \):

\[
A = \max(r_1, r_2, r_3), \quad C = \min(r_1, r_2, r_3)
\]

The corresponding directions of these axes are given by the columns of the matrix \( V \), which can be used for alignment studies or orientation tracking.

\subsection{Log-likelihood calculations accounting for correlated projections}

For each redshift and stellar-mass subset, we fit a two-dimensional kernel density estimate (KDE) 
to the observed $(B/A,\,C/A)$ distribution (using Scott's rule for the bandwidth; we verified robustness 
to bandwidth choices). Given a set of simulated galaxies $\{(x_i, y_i)\}_{i=1}^N$ from either CDM or WDM, 
we compute the total log-likelihood under the observational KDE:
\[
\ln \mathcal{L}_{\mathrm{tot}} = \sum_{i=1}^{N} \ln f_{\mathrm{obs}}(x_i, y_i),
\]
where $f_{\mathrm{obs}}$ is the KDE density. To account for multiple orthogonal projections of the same halo, 
we assign each projection a weight of $1/n_{\mathrm{proj}}$ (three per halo in this case), so that 
$\ln \mathcal{L}_{\mathrm{tot}}$ reflects the weighted sum over projections.
Uncertainties on $\ln \mathcal{L}_{\mathrm{tot}}$ and on $\Delta\ln \mathcal{L}_{\mathrm{tot}}$ (WDM$-$CDM) 
are estimated via bootstrap resampling at the \emph{halo} level (1000 replicates).

\section*{Data availability}

The simulation data used in this work amount to nearly 4~TB in total, with each snapshot divided into multiple large files ($\sim$32~GB per snapshot). Due to their size, storage limitations, and collaboration agreements among the participating institutions, these data cannot be publicly shared. Access to the simulation outputs may be granted upon request, subject to data sharing agreements that ensure appropriate use and approval by all team members and institutions. A complete set of simulated galaxies and 3D animations illustrating the evolution of these simulations can also be provided upon request from the corresponding author.

\section*{Code Availability}
We use the Minimum Volume Enclosing Ellipsoid (MVEE) implementation by Gabriel-p, available at \url{https://gist.github.com/Gabriel-p/4ddd31422a88e7cdf953}. 
For visualization, ellipses are plotted using the \texttt{matplotlib.patches.Ellipse} class, documented at \url{https://matplotlib.org/3.1.1/api/_as_gen/matplotlib.patches.Ellipse.html}.

\section*{Acknowledgements}
 AP, GS and TB are grateful to the DIPC for generous support. AP thanks the Center for Astrophysics | Harvard \& Smithsonian for warm hospitality. RE acknowledges the support from grant numbers 21-atp21-0077, NSF AST-1816420, and HST-GO-16173.001-A as well as the Institute for Theory and Computation at the Center for Astrophysics. We are grateful to the supercomputer facility at Harvard University where most of the simulation work was done. RW acknowledges support from NASA JWST Interdisciplinary Scientist grants NAG5-12460, NNX14AN10G and 80NSSC18K0200 from GSFC. This work has been supported by the Spanish project PID2020-114035GB-100  (MINECO/AEI/FEDER, UE). HNL, Gs and TB are supported by the Collaborative Research Fund under Grant No. C6017-20G which is issued by Research Grants Council of Hong Kong S.A.R. Finally, we are grateful to Lindsay Oldham for her thoughtful guidance and assistance in improving this work during the review process.

\section*{Author contribution}

 AP and TB designed and coordinated the work, prepared the figures, and drafted the manuscript. AP and HNL carried out all necessary simulations, with support from RE, PM, and MV. GS, LH, CC, and RW contributed to the analysis and interpretation of the data and results, and to the final manuscript.

 \section*{Competing  Interest}

The authors declare no competing interests.

\clearpage
\begin{table}[H]
	\centering
\begin{tabular}{|c|c|c|}
\hline
& $\Delta \ln \mathcal{L}_{\mathrm{tot}} $ & $\chi^2$   \\
\hline
 3D Comparison at $3<z<8$ (Fig 2 Top)&  723.69 & - \\
 2D Comparison at $3<z<8$ (Fig 2 Bottom)& 6.24 & - \\
 3D Comparison at $2<z<3$ (Fig 3 Top)& 12.44 &- \\
 2D CEERS Comparison at $2<z<3$ (Fig 3 Middle)& 20.76 & - \\
 
 2D CANDELS Comparison at $2<z<3$ (Fig 3 Bottom)& 18.06 & - \\
 Banana Distribution at $2<z<6$ (Fig 5)& 17.69 & - \\
 $b/a$–Redshift Relation for WDM (Fig 6 Left)& - & 44.11 \\
 $b/a$–Redshift Relation for CDM (Fig 6 Left)& - & 235.36 \\
 $b/a$–Redshift Relation for $\psi$DM (Fig 6 Right)& - & 29.91 \\
 $b/a$–Redshift Relation for WDM (Fig 6 Right)& - & 2.82 \\
 $b/a$–Redshift Relation for CDM (Fig 6 Right)& - & 78.63 \\
 \hline
\end{tabular}
 \caption{ {\bf Log-likelihood comparison between WDM and CDM models.} Presents the log-likelihood differences between warm dark matter (WDM) and cold dark matter (CDM), based on the results shown in the corresponding figures. 
Each entry refers to the figure indicated in the label (e.g., \textit{Figure$_{3\mathrm{D}}$} corresponds to the comparison in three dimensions, \textit{Figure$_{2\mathrm{D}}$} to the two-dimensional projection, and so on). 
In all cases the difference is positive, indicating a systematic preference for WDM over CDM. 
In addition, we report the $\chi^2$ values that quantify the agreement of the different models with the observations ( see Figure \ref{Fig:curve}), providing an independent measure of the quality of the fit in each scenario. \textit{Note: $\Delta \ln \mathcal{L}_{\mathrm{tot}}$ values are raw log-likelihood differences (WDM$-$CDM), not $\Delta\chi^2$; under Gaussian assumptions one has $-2\,\Delta \ln \mathcal{L} \approx \Delta\chi^2$.}}
\label{tab:1}
\end{table}



 \begin{figure}[H]
    \centering
     \includegraphics[width=175mm,height=7.5cm]{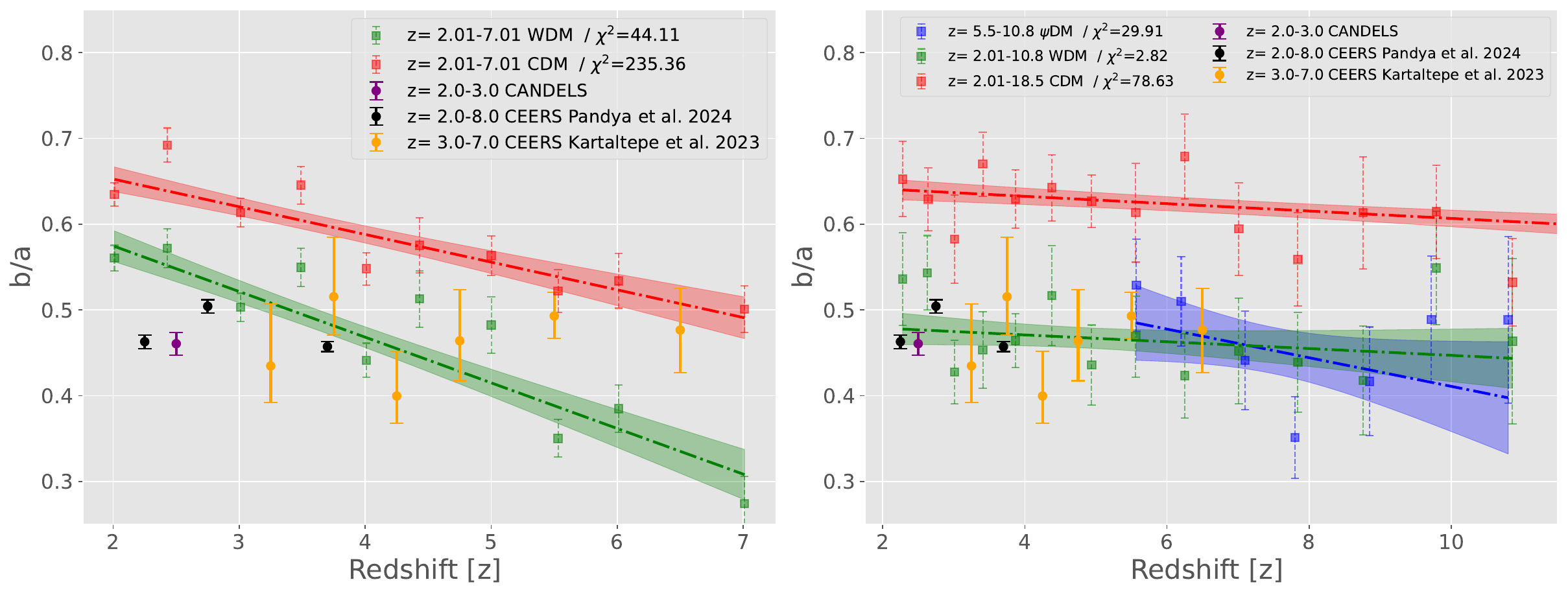}
      \caption{{\bf Redshift dependence of the projected axis ratio $b/a$} compared with CEERS (black \cite{Pandya:2024} and orange \cite{Kartaltepe:2023} points with error bars) and CANDELS (purple \cite{VDW:2014} data points) across the full redshift range covered by our simulations. \textbf{Left panel:} results from the $10\,{\rm Mpc}/h$ and $5\,{\rm Mpc}/h$ simulations, corresponding to virial and stellar mass ranges of $M_{200}=10^{8}$--$10^{12}\,M_{\odot}$ and $M_{\star}=10^{7}$--$5\times10^{10}\,M_{\odot}$, respectively. \textbf{Right panel:} results from the higher-resolution $1.7\,{\rm Mpc}/h$ simulation by Ref\cite{Mocz:2020}, covering halo and stellar mass ranges of $M_{200}=10^{7}$--$5\times10^{10}\,M_{\odot}$ and $M_{\star}=5\times10^{6}$--$5\times10^{8}\,M_{\odot}$, respectively. Error bars denote the $1\sigma$ dispersion around the median values. Note that the higher-redshift CEERS data (last two orange points, $z>5$) correspond to relatively small galaxies for which PSF smoothing may bias the observed $b/a$ toward rounder shapes, possibly leading to an overestimation of the mean values.}

    \label{Fig:curve}
\end{figure}

\newpage

 \begin{center}
    
\textbf{\huge Supplementary Material}

\end{center}

Two higher resolution simulations are analysed in the same way as the main text, to explore the lower galaxy and stellar mass range, appropriate for deeper JWST surveys. Firstly, with a box of 5Mpc$/h$ box, we compare CDM and WDM and secondly a set of smaller boxes, of side  1.7Mpc$/h$, that allows us to include $\psi$DM for comparison with CDM and WDM. It is also important to highlight that this small-box simulation is the original motivation for this work, as it was carried out in 2019\cite{Mocz:2019,Mocz:2020}, several years {\it before} the advent of JWST, and thus represents genuine theoretical priors. This  substantially improves the statistical significance of the main analysis and reinforces our conclusion that WDM and also $\psi$DM lead to more prolate galaxy shapes than CDM. The difference between CDM and WDM in terms of elongation is consistently observed across all stellar masses and scenarios. Moreover, as shown in almost all Figures, the projected semi-axial ratios are consistently lower for WDM and in better agreement with the observational data compared to CDM. In addition, the panels displaying the 3D stellar morphology in Figures 2 and 3 and Supplementary Figures  2 and 5 demonstrate that WDM and $\psi$DM produce a larger number of prolate galaxies over the full range of stellar mass.

The marked difference in predicted stellar morphology that we have highlighted between CDM and the other two DM classes, $\psi$DM \& WDM, is a consequence of the difference in terms of the initial power spectrum and its subsequent evolution. For CDM, the scale-free formation of DM halos extends to arbitrarily lower halo masses, with higher DM concentrations. In contrast, the small-scale suppression of the power spectrum is inherent to both $\psi$DM and WDM. In the case of $\psi$DM, this limits the minimum scale of structure to below the de Broglie wavelength set by the boson mass. Consequently, DM subhalos below $\simeq 10^{9} M_{\odot}$ are suppressed for a boson mass of $\simeq 10^{-22}$ eV and motivated by the observed $\simeq 0.3$ kpc scale of dwarf galaxy cores \cite{Chan:2015,Pozo:2023}. For WDM it is the early relativistic free streaming that smooths the density field on small scales, which also means that low mass galaxy formation is strongly suppressed but in a physically very different way than $\psi$DM which is always non-relativistic, corresponding to a KeV scale particle mass for WDM to approximately match the Kpc size cores of dwarf galaxies.

This suppression of small scale structure in both $\psi$DM and WDM translates into a significant delay in the onset of galaxy formation relative to CDM, which is very evident in the simulations by Ref\cite{Mocz:2020} (Right panel of Figure 5). Furthermore, the simulations predict the development of long, smooth filaments forming first in both WDM and $\psi$DM,  ahead of galaxy formation; see Figure 3 of Ref\cite{Mocz:2020}. This contrasts with the fragmented filaments of CDM, where low-mass subhaloes quickly form due to the presence of small scale power. In contrast, for $\psi$DM \& WDM, fragmentation is not seen along the filaments in the simulations, reflecting the small-scale cutoff of the power spectrum so that dark matter is distributed smoothly along these filaments. In addition to the shape difference between the elongated galaxies of $\psi$DM \& WDM and the irregular spheroidal of CDM, we notice that the stellar profiles are more concentrated compared to $\psi$DM \& WDM, for the corresponding galaxies seen in all three simulations, due to the earlier star formation with CDM and the absence of infall from filaments.

It is important to remember that the data points shown in the different panels of Figures 2,  3 and 4 correspond to different projections of 2 to 25 individual galaxies throughout their cosmological evolution. All figures in the main body show results from the primary simulation, except Figure 5, which also includes results from two additional simulations presented in the supplementary material for statistical robustness. The first repeats the simulations from the main body but in a box of half the size (5 Mpc$/h$), in order to increase the resolution of the analyzed galaxies. This leads to better-resolved measurements of the galaxies' ellipsoidal axes and extends the analysis to lower stellar masses, below $10^9\,M_{\odot}$. While the reduced box size limits the number of halos, it still provides up to 15 matched galaxies, depending on the redshift of each snapshot. Finally, the supplementary material also includes an analysis of the simulations carried out in Ref\cite{Mocz:2020}, which motivated this work and represent genuine priors to the JWST observations. These simulations allow us to include the $\psi$DM scenario for direct comparison with WDM and CDM. The box size in these simulations is much smaller, at 1.7 Mpc$/h$, and $\psi$DM results are only available for $z > 5.5$ due to the resolution limits imposed by the de Broglie wavelength. However, results for WDM and CDM extend down to $z = 2.01$. Despite the limited volume, this simulation yields the most accurate galaxy shape measurements among the three scenarios, making it an ideal testbed for evaluating $b/a$ proportions. Including different simulations increases the amount of data and allows us to verify that, across the three scenarios we have run, the results are consistent. It is important to clarify that the data points shown in the different panels of Supplementary Figure 4 correspond to different projections of three individual galaxies throughout their cosmological evolution. We acknowledge that using such a small number of galaxies may affect the statistical significance of the results. However, we were constrained by the limitations of the $\psi$DM simulations—specifically, the small size of the simulation box, which resulted in a redshift and stellar mass mismatch with the data. It is important to remember that this simulation \cite{Mocz:2019} was done before the analyzed observations and was not done to answer the topic of this work, but it is worthwhile to include the results of $\psi$DM in order to assess its behavior and compare its compatibility with that of WDM/CDM.

\section{Below $10^9$ $ M_{\odot}$ stellar masses}

This appendix describes larger hydro-simulations feasible for WDM and CDM (but not $\psi$DM),  that extend to lower stellar masses, ($10^7-10^9 M_{\odot}$) to allow an additional comparison with the CANDELS and CEERS surveys at $z>2$, where resolution limitations arise for the main body simulation. These extra simulations have a smaller box size of $5.1 \, h^{-1} \, \mathrm{Mpc}$, with a $512^3$ resolution used for dark matter and baryon gas particles, to give a mass resolution of $1.14 \times 10^{4} \, M_{\odot}$, making the shapes of the galaxies more detailed and reliable than in the main body simulation. The simulation evolves from redshift $z = 127$ to $z = 1.01$, adopting cosmological parameters measured by the \textit{Planck} satellite\cite{Planck:2016}, with $\sigma_8 = 0.8$, in agreement with the value recommended by Ref\cite{Naoz:2012}. The revised setup yields halos with virial masses in the range $10^8$–$10^{11}\,M_{\odot}$, but with stellar masses that do not match the data ($10^7$–$5\times 10^{9}\,M_{\odot}$), as most halos in this setup do not reach $10^9\,M_{\odot}$. However, those within the redshift range $2 < z < 3$ do exceed $10^8\,M_{\odot}$.
 Additionally, this setup produces 15 matched galaxies across the two dark matter models. This substantially improves the statistical significance of the main analysis and reinforces the robustness of our conclusions. 

\begin{figure*}[!htbp]
    \centering
    
 \includegraphics[width=1\textwidth,height=7.5cm]{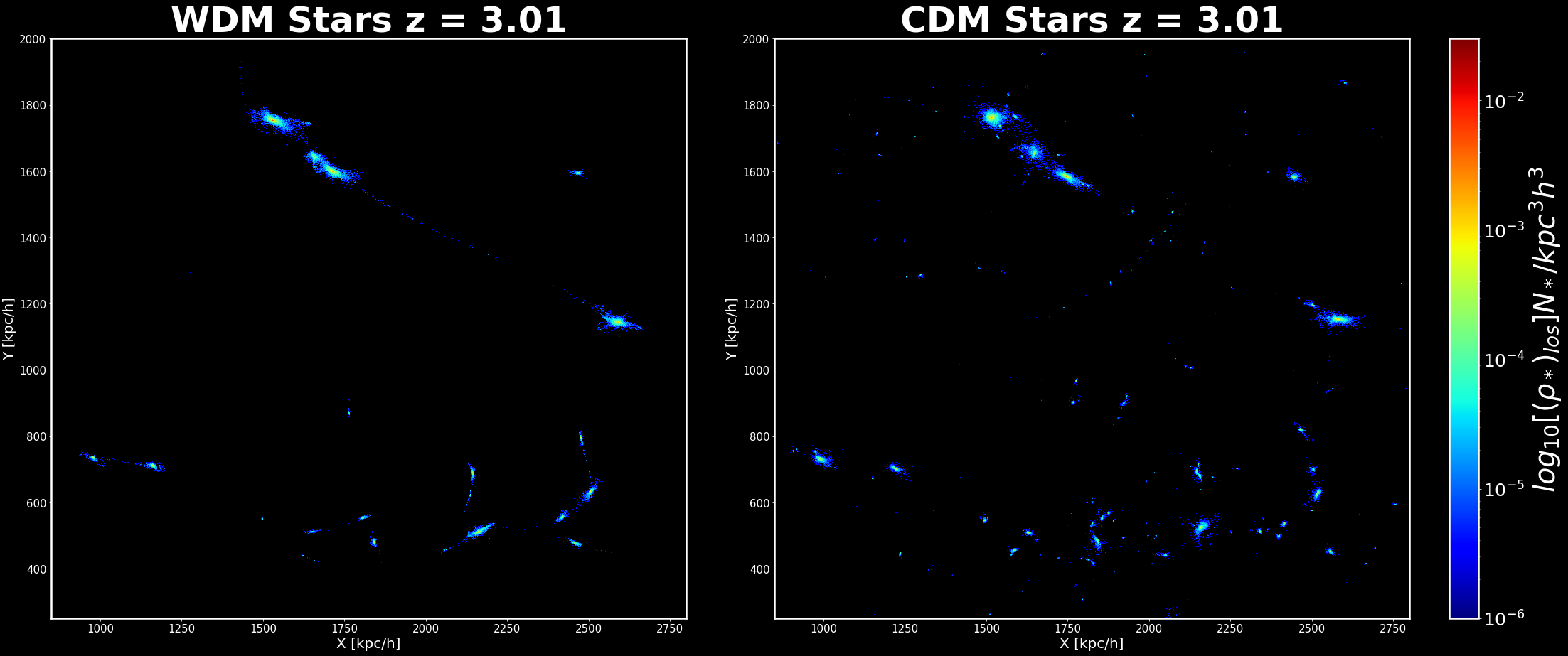}

    \caption*{ {\ \bf Supplementary Figure 1. Zoomed view of galaxy stellar shapes.} This illustrates that WDM generated galaxies shown on the left are noticeably more elongated than the corresponding galaxies 
    of CDM on the right and with smoother filaments that are responsible the generally more prolate galaxy shapes formed with WDM.  }

    \label{Fig:zoom}
\end{figure*}

We zoom in on a subset of these newly simulated galaxies in Supplementary Figure 1 at the mean redshift of $z=3.01$, to illustrate the visible difference between the stellar profiles of galaxies between WDM and CDM. For WDM each galaxy is noticeably elongated, reflecting the evident filament structure visible for WDM. By comparison, each CDM galaxy is noticeably rounder and less elongated by comparison with the corresponding WDM galaxy, and without associated filaments. The statical distribution of these WDM and CDM galaxy shapes traced by the stars is compared with the CANDELS survey which spans $2<z<3$ (bottom row Supplementary Figure 2) and also for CEERS in this low redshift range, to complement the limited resolution of the main body simulation at this redshift range and to extent the analysis to lower stellar mass halos. The stellar masses of the galaxies formed in this relatively small box are generally lower than those typical of the CEERS and CANDELS surveys. These galaxies may be more suitable for comparison with deeper, forthcoming JWST surveys that probe lower stellar masses. Many more lower-mass galaxies are produced by CDM than WDM, as can be seen in Supplementary Figure 1, but only the relatively massive galaxies are in common with WDM, as shown in Supplementary Figure 2. These comparisons reinforce the tendency we identified at higher redshift in the main text, with the maximum of the observed distribution more densely populated by WDM simulated galaxies, and that CDM shows a tendency towards smaller projected elongation, i.e. larger b/a than the data on average when considering both comparison shown here with CANDELS and CEERS. Finally in Supplementary Figure 3 we also compare the new simulations with the "banana" distribution introduced by Ref\cite{Pandya:2024} for a purely prolate distribution, using the galaxies in common between WDM and CDM, for which $M_{200}>10^{9}M_{\odot}$ similar to the relatively massive galaxies comparison the CEERS and CANDELS selection functions. We see the simulated WDM galaxies are concentrated near the predicted peak density for the prolate population at $b/a \simeq 0.3$ and with a narrower range major axis distribution than for CDM for which the mean b/a is larger.

    
   


\begin{figure*}[!htbp]
    \centering
\includegraphics[width=1\textwidth,height=17.5cm]{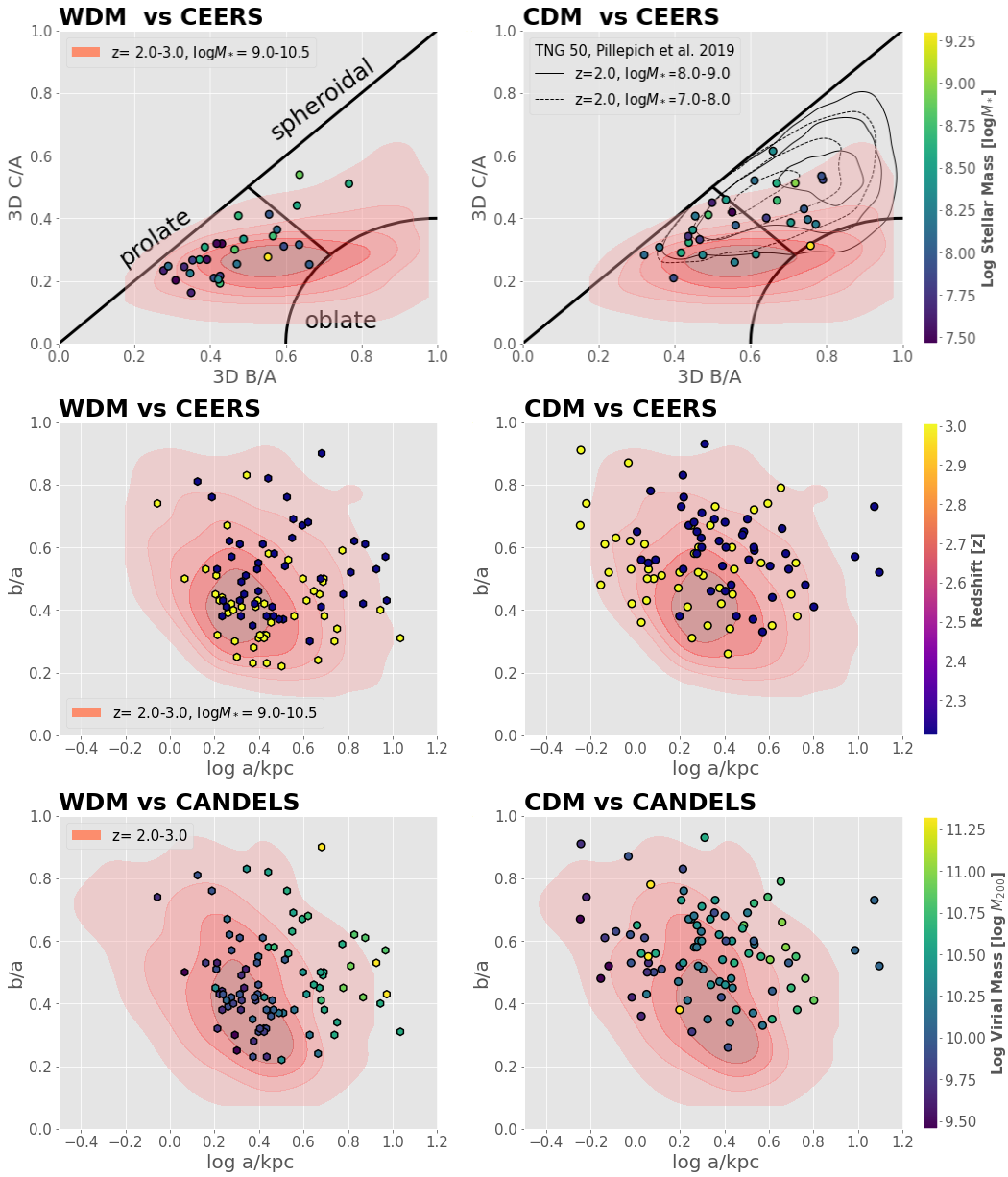}

    \caption*{ {\bf \bf Supplementary Figure 2. Stellar morphology of the WDM and CDM simulation at low redshift $2<z<3$}. The points shown here are for the simulated galaxies in the new larger volume simulations of WDM and CDM  compared to contours representing the observed CEERS and CANDELS surveys. \textbf{Top row:} The black boundaries classify 3D ellipsoids as oblate, spheroidal, or prolate, following the criteria of Refs\cite{Wel:2014,Zhang:2019}. The red shaded area represents the distribution of measurements based on JWST data from Ref\cite{Pandya:2024} for CEERS galaxies at low redshift, $z = 2-3$ ($\Delta \ln \mathcal{L}_{\mathrm{tot}} = 6.61$ in favor of WDM.). Each point in the figure represents a galaxy, although in some cases the same galaxy is shown at multiple redshifts. In total, the sample includes 15 distinct galaxies at $z = 3$ and up to 16 at $z = 2$. The solid and dashed black contours represent the expected distribution of CDM galaxies from TNG50\cite{Pillepich:2019}. Note how our CDM results agree well with the TNG50 predictions for the corresponding stellar masses and redshift range.  \textbf{Middle row:} Similar to the top row, where the red contours now represent the CEERS survey galaxies with $z = 2-3$ \cite{Pandya:2024} ($\Delta \ln \mathcal{L}_{\mathrm{tot}} = 8.43$ in favor of WDM.). In both cases WDM galaxies peaks at $b/a \simeq 0.35$ like the data, whereas CDM favoure generally larger b/a. \textbf{Bottom row:} Projected semi-axis ratio ($b/a$) vs. projected semi-major axis ($a$) for comparison with the red shaded area for $z = 2-3$ of the CANDELS survey \cite{VDW:2014}. The coloured data points correspond to independently calculated ellipticities of galaxies in our simulations, based on the three orthogonal projections of each galaxy. The sample includes projections of 15 halos at $z = 3$ and up to 16 halos at $z = 2$ ($\Delta \ln \mathcal{L}_{\mathrm{tot}} = 6.86$ in favor of WDM.). 
 }
    \label{Fig:compnewsimu}
\end{figure*}

\begin{figure*}[!htbp]
    \centering

    \includegraphics[width=1\textwidth,height=6.0cm]{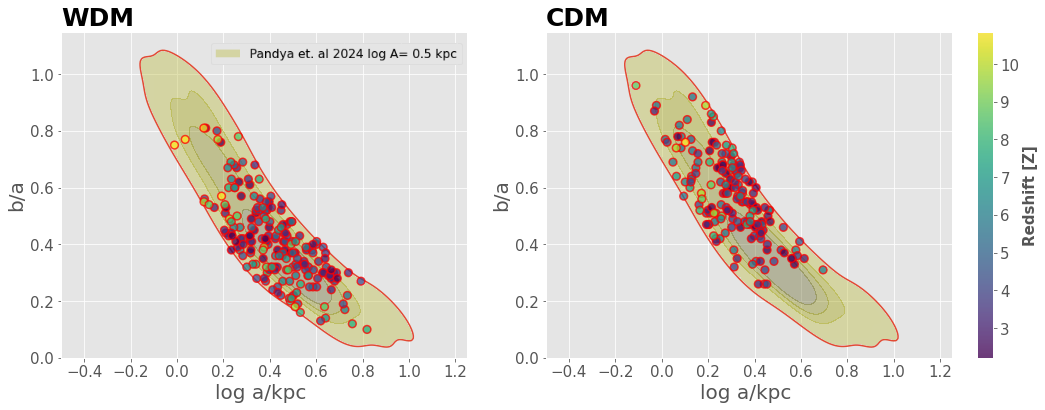}

    \caption*{ {\bf \bf Supplementary Figure 3. Projected axis ratio, b/a, vs. semi-major axis.} 
    The data points represent the resulting $b/a$ values for the halos from the simulations of the box size of 5Mpc/h and the simulations of the 1.7Mpc/h by Ref\cite{Mocz:2020} covering the full redshift range $z = 2-5$. These "banana" contours represent a purely prolate population of triaxial ellipsoids in projection, corresponding to a fixed 3D major axis of $\log (A/\text{kpc}) = 0.5$ with a Gaussian spread of 0.03 dex \cite{Pandya:2024}.The peak of the projected distribution for WDM is in significantly better agreement with the density peak of the “banana” than for CDM for which the mean b/a is larger and less triaxial than WDM, $\Delta \ln \mathcal{L}_{\mathrm{tot}} = 60.21$ in favor of WDM.}
    \label{Fig:banana2}
\end{figure*}

\newpage
\section{Wave Dark Matter analysis}

In addition to these traditional DM classes, we explore ``Wave Dark Matter",  $\psi$DM, which is proving increasingly viable as a non-relativistic dark matter candidate. This is described as a coherent condensate comprised of de Broglie scale waves, including a prominent standing wave ground state, or soliton, near the center of each virialised halo \cite{Schive:2014,Veltmaat:2018,Veltmaat:2019,Mocz:2017,Chan:2022,Pozo:2020}. Initially termed ``Fuzzy" for quantum uncertainty\cite{Hu:2000}, it is now clear that ``Wave" is a more characteristic description \cite{Schive:2014,Hui:2021b}, as revealed by the first simulations where pervasive waves fully modulate the boson density everywhere on the de Broglie scale, ranging from constructive to destructive interference that we term $\psi$DM. To date a number of relatively small volumes have been simulated for $\psi$DM and these are limited to high redshift, as unlike N-body based simulations of CDM and WDM, the de Broglie waves of $\psi$DM must be followed on a 3D grid with sufficient spatial and temporal resolution to accurately predict evolution, becoming more difficult at lower redshift as the box size increases relative to the de Broglie scale that needs to be followed. The details of this wave behaviour is now established in independent simulations utilising different grid based methods \cite{Schive:2014, Mocz:2017, Li:2021,Ivan}. Here we examine the comprehensive hydrodynamical simulations performed by Ref\cite{Mocz:2020}, which remain unique in following the co-evolution of the $\psi$DM waves together with gas hydrodynamics and a self-consistent model for star formation along with comparison simulations of CDM and WDM starting from the same initial conditions \cite{Mocz:2019,Mocz:2020}. These simulations incorporate sub-grid modelling of relevant feedback processes, including SNII driven winds tied to the predicted star formation, following Ref\cite{Vogelsberger:2013}, as outlined in Ref\cite{Mocz:2020}.

\begin{figure*}
    \centering
\includegraphics[width=1\textwidth,height=12.0cm]{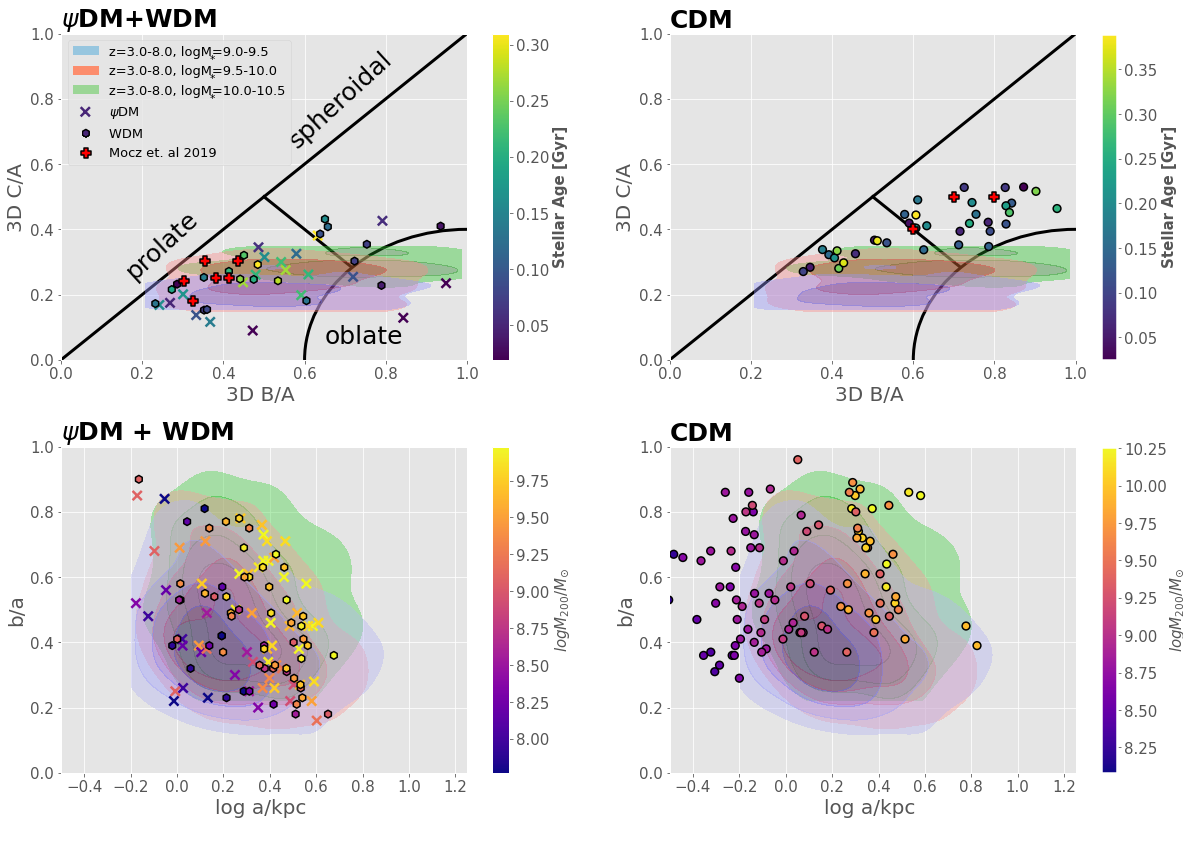}

    \caption*{ {\bf \bf Supplementary Figure 4. 2D \& 3D stellar morphology of simulated galaxies compared with observations}. All time frames are included and the color scale indicates stellar age (top) and galaxy virial mass $M_{200}$ (bottom), with overlapping points offset slightly for clarity. The left panels combine $\psi$DM and WDM to improve the statistics as there is no noticeable difference between these two classes of DM. The right panels show galaxy shapes predicted for CDM. {\bf Top row:}  The black boundaries define 3D ellipsoids as oblate, spheroidal, or prolate, following Refs\cite{Wel:2014,Zhang:2019}. Colour scales indicate the stellar number weighted age of the stars in each simulated galaxy, determined at each time frame of the simulations. The shaded areas represent the distribution of JWST based measurements from Ref\cite{Pandya:2024}. Red data points represent independently calculated ellipticities of the galaxies by Ref\cite{Mocz:2020} for these simulated galaxies. Each point in the figure represents a galaxy, although in some cases the same galaxy is shown at multiple redshifts. In total, the sample includes three distinct galaxies. $\Delta \ln \mathcal{L}_{\mathrm{tot}} = 176.82$ in favor of WDM/$\psi$DM. {\bf Bottom row:} Projected semi-axis ratio, $b/a$, vs. projected semi-major axis $a$, for comparison with the observations represented by shaded areas for z $>$3 \cite{Pandya:2024}. This shows the larger spread towards smaller b/a with WDM \& $\psi$DM, similar to the data when the simulated galaxies are young and lower mass, as indicated by the color bar.
    The data points correspond to independently calculated ellipticities of galaxies in our simulations, based on the three orthogonal projections of each galaxy. The sample includes projections for the three galaxies throughout all the simulated redshifts. $\Delta \ln \mathcal{L}_{\mathrm{tot}} = 107.59$ in favor of WDM/$\psi$DM. over CDM, reinforcing the visibly evident discrepancy with CDM. A full set of simulated galaxies and 3D animations showing the evolution of the three comparison DM model simulations can be obtained by request. }
    \label{Fig:comp}
\end{figure*}

We also emphasize that simulations analysed here were made several years before the advent of JWST and hence represent genuine priors. In this section, we will highlight the obvious similarity of the WDM and $\psi$DM predictions in terms of the formation and the evolution of galaxy morphology. Note, the overlap in redshift with the simulations is set by $\psi$DM where the lowest feasible redshift is $z=5.56$, given the grid based limiting resolution of the joint gas+$\psi$DM simulation. We will emphasize that for this specific simulation, age since formation, rather than redshift, is most relevant in comparison of the data with the simulations, and most of the observed galaxies at $z>3$ are likely on average to be younger than the simulated galaxies at fixed redshift, as the simulation volume has been designed to be sufficiently overdense initially, such that a sample of galaxies is generated relatively early, well above the limiting lower redshift, $z>5.56$ of the $\psi$DM simulation.

\begin{figure*}
    \centering
    \includegraphics[width=1\textwidth,height=17cm]{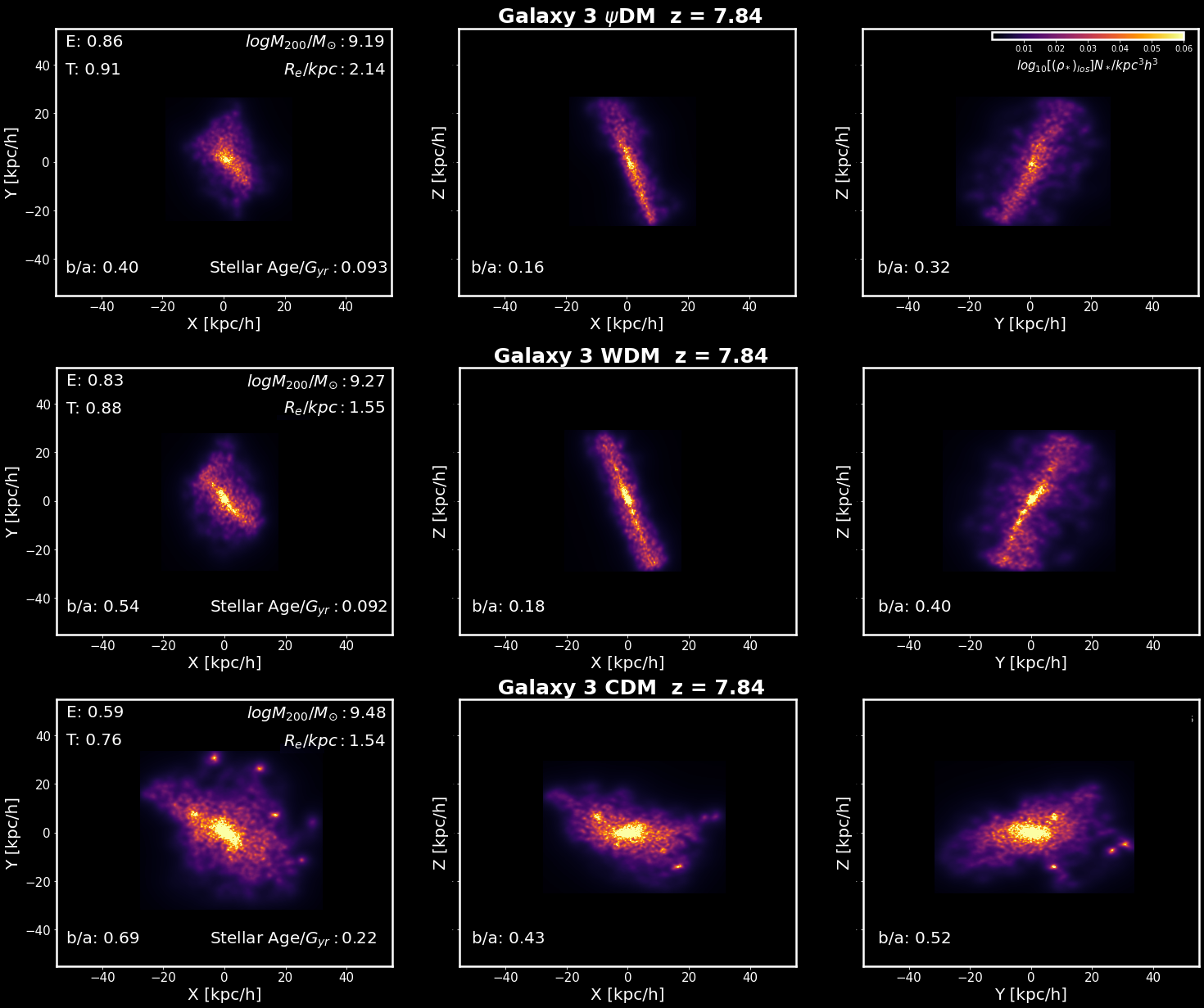}
   
    \caption*{\textbf{ \bf Supplementary Figure 5. Projections in 3D of a representative simulated galaxy.} This comparison illustrates how readily we can differentiate between the prolate shape of galaxies formed in $\psi$DM and WDM from the more spheroidal 3D stellar morphology predicted for the identically located galaxy within the CDM simulation. Simulation scales are shown in comoving units. It is also clear that for CDM there are several prominent subhalos. This difference in shape presented in this picture persists over time. Notice that for CDM, several visible subhalos are also predicted due to early merging, but these are uncommon for $\psi$DM and WDM where early merging is rare, following from the suppression of power at high-k. The row headers indicate the halo number, the DM model, the redshift of the data, and the estimated projected  $b/a$ for each projection. E and T are ellipticity and triaxiality respectively, where a spheroidal ellipsoid would have $E \sim 0$, whereas oblate and prolate ellipsoids would have $E \sim 1$. On the other hand, oblate ellipsoids have $B \sim A$ which means $T \sim 0$, and prolate ellipsoids have $B \sim C$ so $T \sim 1$ (see Ref\cite{Pandya:2024} for a deeper explanation). A full set of simulated galaxies and 3D animations showing the evolution of the three comparison DM model simulations can be obtained by request.
}
    \label{elongated}
\end{figure*}

The bulk of observed CEERS galaxies lie well above the PSF limit, as shown in Figure 4 of Pandya, with $2r_e > $ PSF-FWHM and similar in size to our predictions of $\psi$DM $\&$ WDM, whereas a large fraction of galaxies predicted for CDM are relatively small, comparable with the PSF and further work will be required to folds in the PSF and pixel binning for an accurate comparison. For the data–simulation comparison shown in Supplementary Figure 4, we computed the total log-likelihood ($\ln \mathcal{L}_{\mathrm{tot}}$) of $\psi$DM, WDM, and CDM with respect to the observational dataset, using the same metric and methodology described in the Methods section. The  $\Delta\ln \mathcal{L}_{\mathrm{tot}}$ strongly favoring $\psi$DM and WDM  over CDM. Such large differences correspond to overwhelming statistical evidence against CDM, consistent with the visible discrepancy in the figure, where approximately one-third of the simulated CDM points lie outside the observed distribution and those that overlap tend to favor larger ellipticities than observed.

We also present the three projections of a representative simulated galaxy from Ref\cite{Mocz:2020}, where the resolution is maximun, in Supplementary Figure 5 to illustrate how the 3D appearance readily differentiates between the prolate shape of galaxies formed in $\psi$DM \& WDM from the more spheroidal 3D stellar morphology predicted for the ``same" galaxy with CDM (i.e. identified at the same spatial location in the three DM models simulations). It is also clear that for CDM there are several prominent sub-halos visible in the stellar distribution in this relatively young galaxy. This is a common feature of all the young galaxies $<1$Gyr in age seen in the CDM simulation and is an important distinction as these subhalos are not present for WDM \& $\psi$DM because of the suppression of low mass halos in the power spectrum for both these classes of DM. This behavior may be seen for the full simulations published in Ref\cite{Mocz:2020} and 3D videos showing the evolution of the three comparison DM model simulations can be obtained by request.

 \begin{center}
    
\textbf{\huge Extended Data}


\begin{figure}[H]
    \centering
 \includegraphics[width=0.65\textwidth,height=9cm]{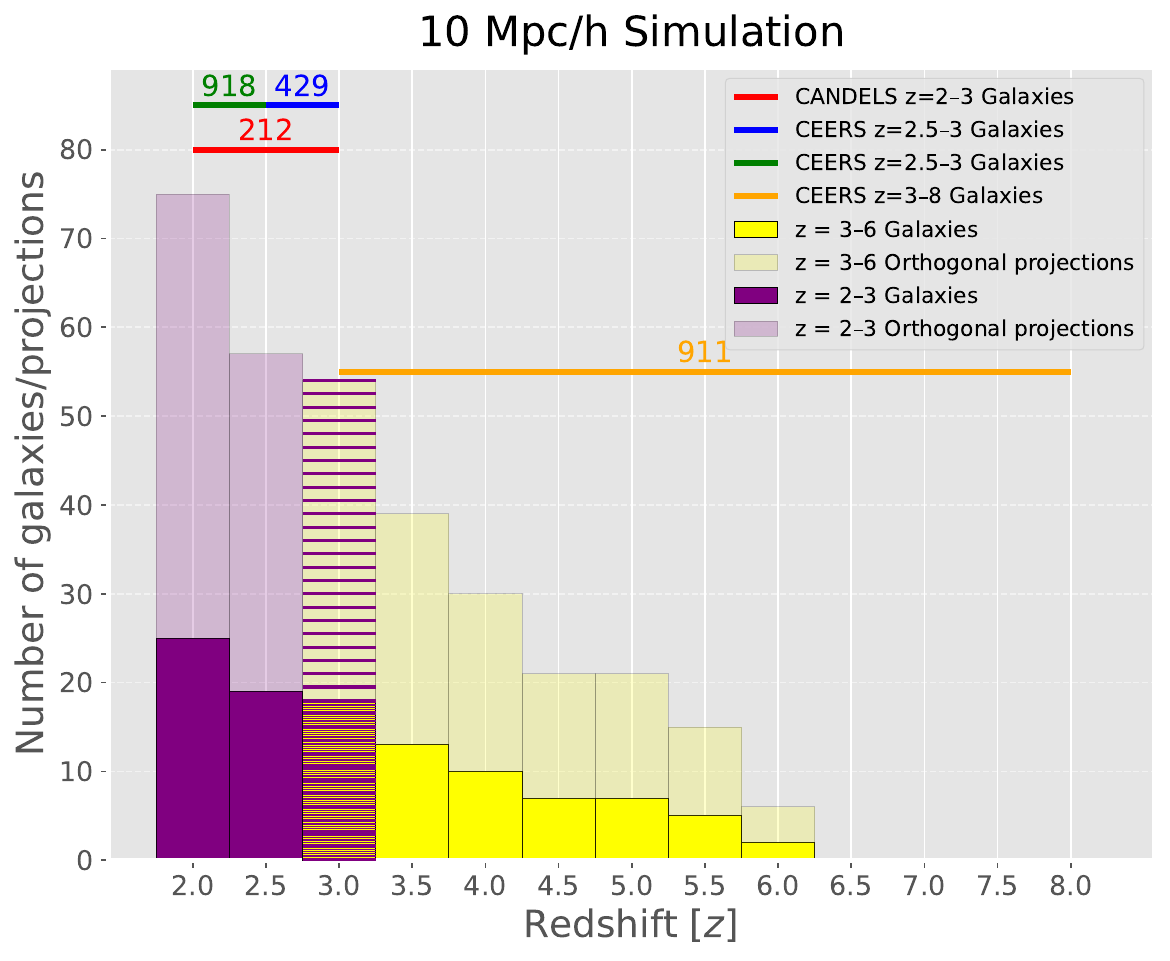}
 \caption*{ {\ \bf Extended Data Figure 1. Number of galaxies resulting from the 10~Mpc/$h$ box simulation}. We present the number of distinct individual halos resulting from the 10~Mpc/$h$ simulation described in the main body of the work. The transparent bars indicate the number of orthogonal projections used in the comparison with the data from Ref\cite{Pandya:2024,VDW:2014}, with three projections per halo, each representing a single data point in Figures~\ref{Fig:compnew}, \ref{Fig:compnew2}, and \ref{Fig:banana1}. Yellow bars correspond to the range of halos with $3 < z < 6$ used in Figure~\ref{Fig:compnew}, while purple bars correspond to the halos used in Figure~\ref{Fig:compnew2} for $z < 3$. Note that both colors are used for the $z = 3$ halos, as they are included in both comparisons. The horizontal color bands indicate the total number of galaxies in the respective catalog within each redshift range.}
   \label{Fig:projections}
\end{figure}

\end{center}


\begin{thebibliography}{99}

\bibitem{Fink}
Finkelstein, S.~L. \textit{et~al.}
The Complete CEERS Early Universe Galaxy Sample: A surprisingly slow evolution of the space density of bright galaxies at $z \sim 8.5$--$14.5$.
\textit{Astrophys.\ J.\ Lett.} \textbf{969}, L2 (2024).

\bibitem{Oguri}
Harikane, Y., Ouchi, M. \& Oguri, M. \textit{et~al.}
A comprehensive study of galaxies at $z \sim 9$--$16$ found in the early JWST data: Ultraviolet luminosity functions and cosmic star formation history at the pre-reionization epoch.
\textit{Astrophys.\ J.\ Suppl.} \textbf{265}, 5 (2023).

\bibitem{LF3}
Robertson, B.~E. \textit{et~al.}
Earliest galaxies in the JADES Origins Field: Luminosity function and cosmic star formation rate density 300 Myr after the Big Bang.
\textit{Astrophys.\ J.} \textbf{970}, 31 (2024).


\bibitem{Nathan}
Adams, N.~J. \textit{et~al.}
EPOCHS. II. The ultraviolet luminosity function from $7.5 < z < 13.5$ using 180~arcmin$^{2}$ of deep, blank fields from the PEARLS survey and public JWST data.
\textit{Astrophys.\ J.} \textbf{965}, 169 (2024).

\bibitem{hz1}
Curtis-Lake, E. \textit{et~al.}
Spectroscopic confirmation of four metal-poor galaxies at $z = 10.3$--$13.2$.
\textit{Nat. Astron.} \textbf{7}, 622--632 (2023).


\bibitem{hz2}
D'Eugenio, F. \textit{et~al.}
JADES: Carbon enrichment 350 Myr after the Big Bang.
\textit{Astron.\ Astrophys.} \textbf{689}, A152 (2024).

\bibitem{hz3}
Witstok, J. \textit{et~al.}
Witnessing the onset of reionization through Lyman-$\alpha$ emission at redshift~13.
\textit{Nature} \textbf{639}, 897--901 (2025).

\bibitem{hz4}
Wu, Z. \textit{et~al.}
JADES-GS-z14-1: A compact, faint galaxy at $z \approx 14$ with weak metal lines from extremely deep JWST MIRI, NIRCam, and NIRSpec observations.
\textit{arXiv e-prints}, arXiv:2507.22858 (2025).

\bibitem{Labbe:2023}
Labb{\'e}, I. \textit{et~al.}
A population of red candidate massive galaxies 600~Myr after the Big Bang.
\textit{Nature} \textbf{616}, 266--269 (2023).

\bibitem{Mathee}
Matthee, J. \textit{et~al.}
Little Red Dots: An abundant population of faint active galactic nuclei at $z \sim 5$ revealed by the EIGER and FRESCO JWST surveys.
\textit{Astrophys.\ J.} \textbf{963}, 129 (2024).

\bibitem{Donnan}
Donnan, C.~T. \textit{et~al.}
The evolution of the galaxy UV luminosity function at redshifts $z \simeq 8$--15 from deep JWST and ground-based near-infrared imaging.
\textit{Mon.\ Not.\ R.\ Astron.\ Soc.} \textbf{518}, 6011--6040 (2023).

\bibitem{Kocevski}
Kocevski, D.~D. \textit{et~al.}
The rise of faint, red active galactic nuclei at $z > 4$: A sample of Little Red Dots in the JWST Extragalactic Legacy Fields.
\textit{Astrophys.\ J.} \textbf{986}, 126 (2025).

\bibitem{Kokorev}
Kokorev, V. \textit{et~al.}
A census of photometrically selected Little Red Dots at $4 < z < 9$ in JWST blank fields.
\textit{Astrophys.\ J.} \textbf{968}, 38 (2024).


\bibitem{Wil_dot}
Williams, C.~C. \textit{et~al.}
The galaxies missed by Hubble and ALMA: The contribution of extremely red galaxies to the cosmic census at $3 < z < 8$.
\textit{Astrophys.\ J.} \textbf{968}, 34 (2024).

\bibitem{Epoch}
Carranza-Escudero, M. \textit{et~al.}
Lonely Little Red Dots: Challenges to the active galactic nucleus nature of Little Red Dots through their clustering and spectral energy distributions.
\textit{Astrophys.\ J.\ Lett.} \textbf{989}, L50 (2025).

\bibitem{Naidu}
Naidu, R.~P. \textit{et~al.}
Two remarkably luminous galaxy candidates at $z \approx 10$--12 revealed by JWST.
\textit{Astrophys.\ J.\ Lett.} \textbf{940}, L14 (2022).

\bibitem{Goulding}
Goulding, A.~D. \textit{et~al.}
UNCOVER: The growth of the first massive black holes from JWST/NIRSpec-spectroscopic redshift confirmation of an X-ray luminous AGN at $z = 10.1$.
\textit{Astrophys.\ J.\ Lett.} \textbf{955}, L24 (2023).

\bibitem{Geris}
Geris, S. \textit{et~al.}
JADES reveals a large population of low-mass black holes at high redshift.
\textit{arXiv e-prints}, arXiv:2506.22147 (2025).

\bibitem{RotationJWST}
Pascalau, R.~G. \textit{et~al.}
When relics were made: vigorous stellar rotation and low dark matter content in the massive ultra-compact galaxy GS-9209 at $z = 4.66$.
\textit{arXiv e-prints}, arXiv:2505.06349 (2025).


\bibitem{DE}
D'Eugenio, F. \textit{et~al.}
A fast-rotator post-starburst galaxy quenched by supermassive black-hole feedback at $z = 3$.
\textit{Nature Astron.} \textbf{8}, 1443--1456 (2024).

\bibitem{spiral2}
Wang, W. \textit{et~al.}
A giant disk galaxy two billion years after the Big Bang.
\textit{Nat. Astron.} \textbf{9}, 710--719 (2025).

\bibitem{Exception1}
Xiao, M. \textit{et~al.}
PANORAMIC: Discovery of an ultra-massive grand-design spiral galaxy at $z \sim 5.2$.
\textit{Astron.\ Astrophys.} \textbf{696}, A156 (2025).


\bibitem{Bar1}
G{\'e}ron, T. \textit{et~al.}
Galaxy Zoo CEERS: Bar fractions up to $z \sim 4.0$.
\textit{Astrophys.\ J.} \textbf{987}, 74 (2025).


\bibitem{Bar2}
Huang, S. \textit{et~al.}
Large gas inflow driven by a matured galactic bar in the early Universe.
\textit{Nature} \textbf{641}, 861--865 (2025).


\bibitem{Car1}
Carnall, A.~C. \textit{et~al.}
A massive quiescent galaxy at redshift $z = 4.658$.
\textit{Nature} \textbf{619}, 716--719 (2023).

\bibitem{Exception0}
Glazebrook, K. \textit{et~al.}
A massive, quiescent galaxy at a redshift of $z = 3.717$.
\textit{Nature} \textbf{544}, 71--74 (2017).

\bibitem{exception3}
Glazebrook, K. \textit{et~al.}
A massive galaxy that formed its stars at $z \approx 11$.
\textit{Nature} \textbf{628}, 277--281 (2024).


\bibitem{Chal}
Espejo Salcedo, J.~M. \textit{et~al.}
Galaxy morphologies at cosmic noon with JWST: A foundation for exploring gas transport with bars and spiral arms.
\textit{Astron.\ Astrophys.} \textbf{700}, A42 (2025).

\bibitem{Car2}
Carnall, A.~C. \textit{et~al.}
The JWST EXCELS survey: too much, too young, too fast? Ultra-massive quiescent galaxies at $3 < z < 5$.
\textit{Mon.\ Not.\ R.\ Astron.\ Soc.} \textbf{534}, 325--348 (2024).


\bibitem{Pandya:2024}
Pandya, V. \textit{et~al.}
Galaxies going bananas: Inferring the 3D geometry of high-redshift galaxies with JWST-CEERS.
\textit{Astrophys.\ J.} \textbf{963}, 54 (2024).

\bibitem{Finkelstein:2023}
Finkelstein, S.~L. \textit{et~al.}
CEERS Key Paper. I. An early look into the first 500~Myr of galaxy formation with JWST.
\textit{Astrophys.\ J.} \textbf{946}, L13 (2023).

\bibitem{Excess}
Westcott, L. \textit{et~al.}
EPOCHS. XI. The structure and morphology of galaxies in the epoch of reionization to $z \sim 12.5$.
\textit{Astrophys.\ J.} \textbf{983}, 121 (2025).

\bibitem{Gibson:2024}
Gibson, J.~L. \textit{et~al.}
JADES ultrared flattened objects: Morphologies and spatial gradients in color and stellar populations.
\textit{Astrophys.\ J.} \textbf{974}, 48 (2024).

\bibitem{Ferreira:2022}
Ferreira, L. \textit{et~al.}
Panic! at the Disks: First rest-frame optical observations of galaxy structure at $z > 3$ with JWST in the SMACS~0723 field.
\textit{Astrophys.\ J.} \textbf{938}, L2 (2022).

\bibitem{Cowie:1995}
Cowie, L.~L., Hu, E.~M. \& Songaila, A.
Faintest galaxy morphologies from HST WFPC2 imaging of the Hawaii Survey Fields.
\textit{Astron.\ J.} \textbf{110}, 1576 (1995).

\bibitem{Elmgreen}
Elmegreen, D.~M., Elmegreen, B.~G. \& Sheets, C.~M.
Chain galaxies in the Tadpole Advanced Camera for Surveys field.
\textit{Astrophys.\ J.} \textbf{603}, 74--81 (2004).

\bibitem{Straughn:2006}
Straughn, A.~N. \textit{et~al.}
Tracing galaxy assembly: Tadpole galaxies in the Hubble Ultra Deep Field.
\textit{Astrophys.\ J.} \textbf{639}, 724--730 (2006).

\bibitem{VDW:2014}
van der Wel, A. \textit{et~al.}
3D-HST+CANDELS: The evolution of the galaxy size–mass distribution since $z = 3$.
\textit{Astrophys.\ J.} \textbf{788}, 28 (2014).

\bibitem{Zhang:2019}
Zhang, H. \textit{et~al.}
The evolution of galaxy shapes in CANDELS: from prolate to discy.
\textit{Mon.\ Not.\ R.\ Astron.\ Soc.} \textbf{484}, 5170--5191 (2019).

\bibitem{Law:2012}
Law, D.~R. \textit{et~al.}
An HST/WFC3-IR morphological survey of galaxies at $z = 1.5$–$3.6$. I. Survey description and morphological properties of star-forming galaxies.
\textit{Astrophys.\ J.} \textbf{745}, 85 (2012).

\bibitem{Odewahn:1997}
Odewahn, S.~C., Burstein, D. \& Windhorst, R.~A.  
The axis ratio distribution of local and distant galaxies.  
\textit{Astron.\ J.} \textbf{114}, 2219 (1997).

\bibitem{Rogier1}
Windhorst, R.~A. \textit{et~al.}  
A \textit{Hubble Space Telescope} survey of the mid-ultraviolet morphology of nearby galaxies.  
\textit{Astrophys.\ J.\ Suppl.} \textbf{143}, 113--158 (2002).

\bibitem{DISK1}
Danhaive, A.~L. \textit{et~al.}  
The dawn of disks: unveiling the turbulent ionised gas kinematics of the galaxy population at $z\!\sim\!4$–6 with JWST/NIRCam grism spectroscopy.  
\textit{arXiv e-prints} arXiv:2503.21863 (2025).

\bibitem{DISK3}
Adamo, A. \textit{et~al.}  
The first billion years according to JWST.  
\textit{Nature Astron.} \textbf{9}, 1134--1147 (2025).

\bibitem{DISK4}
Huertas-Company, M. \textit{et~al.}  
COSMOS-Web: the emergence of the Hubble sequence.  
\textit{arXiv e-prints} arXiv:2502.03532 (2025).

\bibitem{Mocz:2020}
Mocz, P. \textit{et~al.}  
Galaxy formation with BECDM – II. Cosmic filaments and first galaxies.  
\textit{Mon.\ Not.\ R.\ Astron.\ Soc.} \textbf{494}, 2027--2044 (2020).

\bibitem{Tomassetti:2016}
Tomassetti, M. \textit{et~al.}  
Evolution of galaxy shapes from prolate to oblate through compaction events.  
\textit{Mon.\ Not.\ R.\ Astron.\ Soc.} \textbf{458}, 4477--4497 (2016).

\bibitem{Pillepich:2019}
Pillepich, A. \textit{et~al.}  
First results from the TNG50 simulation: the evolution of stellar and gaseous discs across cosmic time.  
\textit{Mon.\ Not.\ R.\ Astron.\ Soc.} \textbf{490}, 3196--3233 (2019).

\bibitem{Wel:2014}
van der Wel, A. \textit{et~al.}  
Geometry of star-forming galaxies from SDSS, 3D-HST, and CANDELS.  
\textit{Astrophys.\ J.} \textbf{792}, L6 (2014).

\bibitem{Grogin}
Grogin, N. A. \textit{et~al.}  
CANDELS: The Cosmic Assembly Near-infrared Deep Extragalactic Legacy Survey.  
\textit{Astrophys.\ J.\ Suppl.} \textbf{197}, 35 (2011).

\bibitem{Ceverino:2015}
Ceverino, D., Primack, J. \& Dekel, A.  
Formation of elongated galaxies with low masses at high redshift.  
\textit{Mon.\ Not.\ R.\ Astron.\ Soc.} \textbf{453}, 408--413 (2015).


\bibitem{Kartaltepe:2023}
Kartaltepe, J. S. \textit{et~al.}  
CEERS Key Paper. III. The diversity of galaxy structure and morphology at $z = 3$–$9$ with JWST.  
\textit{Astrophys.\ J.} \textbf{946}, L15 (2023).

\bibitem{GAMA}
Baldry, I. K. \textit{et~al.}  
Galaxy And Mass Assembly: the G02 field, Herschel-ATLAS target selection and data release 3.  
\textit{Mon.\ Not.\ R.\ Astron.\ Soc.} \textbf{474}, 3875--3888 (2018).

\bibitem{FIREBOX}
Klein, C. \textit{et~al.}  
The shape of FIREbox galaxies and a potential tension with low-mass disks.  
\textit{arXiv\ e-prints} arXiv:2503.05612 (2025).

\bibitem{local}
Wang, B., Peng, Y., Cappellari, M., Gao, H. \& Mo, H.  
On the kinematic nature of apparent disks at high redshifts: local counterparts are not dominated by ordered rotation but by tangentially anisotropic random motion.  
\textit{Astrophys.\ J.\ Lett.} \textbf{973}, L29 (2024).

\bibitem{Valle:2023}
del Valle-Espinosa, M. G. \textit{et~al.}  
Spatially resolved chemodynamics of the starburst dwarf galaxy CGCG 007–025: evidence for recent accretion of metal-poor gas.  
\textit{Mon.\ Not.\ R.\ Astron.\ Soc.} \textbf{522}, 2089--2104 (2023).

\bibitem{Schive:2016}
Schive, H.-Y., Chiueh, T., Broadhurst, T. \& Huang, K.-W.  
Contrasting galaxy formation from quantum wave dark matter, $\psi$DM, with $\Lambda$CDM, using \textit{Planck} and \textit{Hubble} data.  
\textit{Astrophys.\ J.} \textbf{818}, 89 (2016).

\bibitem{Duan:2024}
Duan, Q. \textit{et~al.}  
Galaxy mergers in the epoch of reionization – I. A JWST study of pair fractions, merger rates, and stellar mass accretion rates at $z = 4.5$–$11.5$.  
\textit{Mon.\ Not.\ R.\ Astron.\ Soc.} \textbf{540}, 774--805 (2025).

\bibitem{Pusk}
Puskás, D. \textit{et~al.}  
Constraining the major merger history of $z \sim 3$–9 galaxies using JADES: dominant \textit{in situ} star formation.  
\textit{Mon.\ Not.\ R.\ Astron.\ Soc.} \textbf{540}, 2146--2175 (2025).

\bibitem{Windhorst:2024}
Windhorst, R. A. \textit{et~al.}  
Galaxy science with ORCAS: faint star-forming clumps to AB $\leq 31$ mag and $r_{\mathrm{e}} \geq 0.01''$.  
\textit{arxiv e-prints} arXiv:2106.02664 (2021).

\bibitem{heating1}
Brooks, A. M., Kuhlen, M., Zolotov, A. \& Hooper, D.  
A baryonic solution to the missing satellites problem.  
\textit{Astrophys.\ J.} \textbf{765}, 22 (2013).

\bibitem{heating2}
Jeon, S. \textit{et~al.}  
Born to be starless: revisiting the missing satellite problem.  
\textit{Astrophys.\ J.} \textbf{988}, 136 (2025).

\bibitem{Vogelsberger:2013}
Vogelsberger, M. \textit{et~al.}  
A model for cosmological simulations of galaxy formation physics.  
\textit{Mon.\ Not.\ R.\ Astron.\ Soc.} \textbf{436}, 3031–3067 (2013).

\bibitem{Gao:2007}
Gao, L. \& Theuns, T.  
Lighting the Universe with filaments.  
\textit{Science} \textbf{317}, 1527 (2007).

\bibitem{Mocz:2019}
Mocz, P. \textit{et~al.}  
First star-forming structures in fuzzy cosmic filaments.  
\textit{Phys.\ Rev.\ Lett.} \textbf{123}, 141301 (2019).

\bibitem{F2}
Boyarsky, A., Iakubovskyi, D., Ruchayskiy, O., Rudakovskyi, A. \& Valkenburg, W.  
21-cm observations and warm dark matter models.  
\textit{Phys.\ Rev.\ D} \textbf{100}, 123005 (2019).

\bibitem{F3}
Liu, Y., Gao, L., Liao, S. \& Zhu, K.  
Prospects for detecting cosmic filaments in Ly$\alpha$ emission across redshifts $z = 2$–$5$.  
\textit{Astrophys.\ J.} \textbf{984}, 55 (2025).

\bibitem{Ma:2024}
Ma, Z. \textit{et~al.}  
JWST view of three infant galaxies at $z = 8.3$ and implications for reionization.  
\textit{Astrophys.\ J.} \textbf{975}, 15 (2024).

\bibitem{Loiacono:2024}
Loiacono, F. \textit{et~al.}  
A quasar–galaxy merger at $z \sim 6.2$: black hole mass and quasar properties from the NIRSpec spectrum.  
\textit{Astron.\ Astrophys.} \textbf{685}, A121 (2024).


\bibitem{Pandya:2019}
Pandya, V. \textit{et~al.}  
Can intrinsic alignments of elongated low-mass galaxies be used to map the cosmic web at high redshift?  
\textit{Mon.\ Not.\ R.\ Astron.\ Soc.} \textbf{488}, 5580–5593 (2019).

\bibitem{Pozo:2020}
Pozo, A. \textit{et~al.}  
Detection of a universal core–halo transition in dwarf galaxies as predicted by Bose–Einstein dark matter.  
\textit{Phys.\ Rev.\ D} \textbf{110}, 043534 (2024).

\bibitem{Pozo:20232}
Pozo, A. \textit{et~al.}  
Galaxy formation with wave/fuzzy dark matter: the core–halo structure and the solitonic imprint.  
\textit{Astron.\ Astrophys.} \textbf{699}, A308 (2025).

\bibitem{Fudamoto}
Fudamoto, Y. \textit{et~al.}  
Identification of more than 40 gravitationally magnified stars in a galaxy at redshift 0.725.  
\textit{Nat. Astron.} \textbf{9}, 428–437 (2025).

\bibitem{Yan}
Yan, H. \textit{et~al.}  
JWST’s PEARLS: transients in the MACS~J0416.1–2403 field.  
\textit{Astrophys.\ J.\ Suppl.} \textbf{269}, 43 (2023).

\bibitem{Broadhurst:2024}
Broadhurst, T. \textit{et~al.}  
Dark matter distinguished by skewed microlensing in the “Dragon Arc”.  
\textit{Astrophys.\ J.} \textbf{978}, L5 (2025).


\bibitem{Springel:2010}
Springel, V.  
\textit{E pur si muove}: Galilean-invariant cosmological hydrodynamical simulations on a moving mesh.  
\textit{Mon.\ Not.\ R.\ Astron.\ Soc.} \textbf{401}, 791–851 (2010).

\bibitem{Mocz:2017}
Mocz, P. \textit{et~al.}  
Galaxy formation with BECDM – I. Turbulence and relaxation of idealized haloes.  
\textit{Mon.\ Not.\ R.\ Astron.\ Soc.} \textbf{471}, 4559–4570 (2017).

\bibitem{Springel:2015}
Springel, V.  
\textit{N-GenIC: Cosmological structure initial conditions}.  
Astrophysics Source Code Library, ascl:1502.003 (2015).

\bibitem{Lewis:2011}
Lewis, A. \& Challinor, A.  
\textit{CAMB: Code for Anisotropies in the Microwave Background}.  
Astrophysics Source Code Library, ascl:1102.026 (2011).

\bibitem{Hlozek:2015}
Hlozek, R., Grin, D., Marsh, D.~J.~E. \& Ferreira, P.~G.  
A search for ultralight axions using precision cosmological data.  
\textit{Phys.\ Rev.\ D} \textbf{91}, 103512 (2015).


\bibitem{Springel:2018}
Springel, V. \textit{et al.}  
First results from the IllustrisTNG simulations: matter and galaxy clustering.  
\textit{Mon.\ Not.\ R.\ Astron.\ Soc.} \textbf{475}, 676–698 (2018).

\bibitem{feedback}
Jung, M. \textit{et al.}  
The \textit{AGORA} High-resolution Galaxy Simulations Comparison Project. VIII: Disk formation and evolution of simulated Milky Way mass galaxy progenitors at $1<z<5$.  
\textit{arXiv e-prints}, arXiv:2505.05720 (2025).

\bibitem{Pozo:2021}
Pozo, A. \textit{et al.}  
Wave dark matter and ultra-diffuse galaxies.  
\textit{Mon.\ Not.\ R.\ Astron.\ Soc.} \textbf{504}, 2868–2876 (2021).


\bibitem{Aprile:2022}
Aprile, E. \textit{et al.} (XENON Collaboration)  
Search for new physics in electronic recoil data from XENONnT.  
\textit{Phys.\ Rev.\ Lett.} \textbf{129}, 161805 (2022).

\bibitem{Planck:2016}
Planck Collaboration \textit{et al.}  
Planck 2015 results. XIII. Cosmological parameters.  
\textit{Astron.\ Astrophys.} \textbf{594}, A13 (2016).


\bibitem{Naoz:2012}
Naoz, S., Yoshida, N. \& Gnedin, N.~Y.  
Simulations of early baryonic structure formation with stream velocity. I. Halo abundance.  
\textit{Astrophys.\ J.} \textbf{747}, 128 (2012).

\bibitem{Vandenberghe:1998}
Vandenberghe, L., Boyd, S. \& Wu, S.-P.  
Determinant maximization with linear matrix inequality constraints.  
\textit{SIAM J.\ Matrix Anal.\ Appl.} \textbf{19}, (1998).


\bibitem{Chan:2015}
Chan, T.~K., Kere{\v{s}}, D., O{\~n}orbe, J., Hopkins, P.~F., Muratov, A.~L., Faucher-Gigu{\`e}re, C.-A. \& Quataert, E.  
The impact of baryonic physics on the structure of dark matter haloes: the view from the FIRE cosmological simulations.  
\textit{Mon.\ Not.\ R.\ Astron.\ Soc.} \textbf{454}, 2981–3001 (2015).


\bibitem{Pozo:2023}
Pozo, A., Broadhurst, T., Smoot, G.~F., Chiueh, T., Luu, H.~N., Vogelsberger, M. \& Mocz, P.  
Dwarf galaxies united by dark bosons.  
\textit{Phys.\ Rev.\ D} \textbf{109}, 083532 (2024).


\bibitem{Schive:2014}
Schive, H.-Y., Chiueh, T. \& Broadhurst, T.  
Cosmic structure as the quantum interference of a coherent dark wave.  
\textit{Nat.\ Phys.} \textbf{10}, 496–499 (2014).

\bibitem{Veltmaat:2018}
Veltmaat, J., Niemeyer, J.~C. \& Schwabe, B.  
Formation and structure of ultralight bosonic dark matter halos.  
\textit{Phys.\ Rev.\ D} \textbf{98}, 043509 (2018).

\bibitem{Veltmaat:2019}
Veltmaat, J., Schwabe, B. \& Niemeyer, J.~C.  
Baryon-driven growth of solitonic cores in fuzzy dark matter halos.  
\textit{Phys.\ Rev.\ D} \textbf{101}, 083518 (2020).

\bibitem{Chan:2022}
Chan, H.~Y.~J., Ferreira, E.~G.~M., May, S., Hayashi, K. \& Chiba, M.  
The diversity of core–halo structure in the fuzzy dark matter model.  
\textit{Mon.\ Not.\ R.\ Astron.\ Soc.} \textbf{511}, 943–952 (2022).

\bibitem{Hu:2000}
Hu, W., Barkana, R. \& Gruzinov, A.  
Fuzzy Cold Dark Matter: The Wave Properties of Ultralight Particles.  
\textit{Phys.\ Rev.\ Lett.} \textbf{85}, 1158–1161 (2000).


\bibitem{Hui:2021b}
Hui, L.  
Wave dark matter.  
\textit{Annu.\ Rev.\ Astron.\ Astrophys.} \textbf{59}, 247–289 (2021).

\bibitem{Li:2021}
Li, X., Hui, L. \& Yavetz, T.~D.  
Oscillations and random walk of the soliton core in a fuzzy dark matter halo.  
\textit{Phys.\ Rev.\ D} \textbf{103}, 023508 (2021).


\bibitem{Ivan}
Della Monica, R., de Martino, I. \& Broadhurst, T.  
Explaining the oblate morphology of dwarf spheroidals with wave dark matter perturbations.  
\textit{Mon.\ Not.\ R.\ Astron.\ Soc.} \textbf{534}, 2196–2203 (2024).

\bibitem{Vogelsberger:2013}
Vogelsberger, M. \textit{et~al.}  
A model for cosmological simulations of galaxy formation physics.  
\textit{Mon.\ Not.\ R.\ Astron.\ Soc.} \textbf{436}, 3031–3067 (2013).







\end{thebibliography}
\end{document}